\documentclass{emulateapj}
\usepackage{graphicx}
\usepackage{natbib}
\usepackage{amssymb}
\usepackage{float}
\usepackage{makeidx}
\usepackage{amsmath}
\usepackage{subfigure}
\usepackage{subfloat}
\usepackage{captcont}

\nocite{*}

\newcommand{\mmsun}{M_\odot}

\newcommand{\etal}{et~al.\/}

\newcommand{\ovi}{\hbox{O\,{\sc vi}}}

\shortauthors{Werk \etal}
\shorttitle{COS Program LRIS data}

\begin{document} 
\nocite{*}
\slugcomment{Submitted Version Emulate ApJ: August 2011}

\title{The COS-Halos Survey: Keck LRIS  and Magellan MagE Optical Spectroscopy}

\author{Jessica K.\ Werk\altaffilmark{1},
J. Xavier Prochaska\altaffilmark{1},
 Christopher Thom\altaffilmark{2},
  Jason Tumlinson \altaffilmark{2}, 
  Todd M. Tripp \altaffilmark{3}, 
  John M. O'Meara \altaffilmark{4}, 
  Joseph D. Meiring \altaffilmark{3}
}

\altaffiltext{1}{UCO/Lick Observatory; University of California, Santa Cruz, CA $jwerk@ucolick.org$}
\altaffiltext{2}{ Space Telescope Science Institute, 3700 San Martin Drive,  Baltimore, MD}
\altaffiltext{3}{Department of Astronomy, University of Massachusetts, Amherst, MA}
\altaffiltext{4}{Department of Chemistry and Physics, Saint Michael's College, Colchester, VT}

\begin{abstract}
We present high signal-to-noise optical spectra for 67  low-redshift (0.1 $<$ z $<$ 0.4) galaxies that lie within close projected distances (5 kpc $<$ $\rho$ $<$ 150 kpc) of  38 background UV-bright QSOs.  The Keck LRIS and Magellan MagE data presented here are part of a survey that aims to construct a statistically sampled map of the physical state and metallicity of gaseous galaxy halos  using the {\emph{Cosmic Origins Spectrograph}} (COS) on the {\it Hubble Space Telescope} ({\it HST}). We provide a detailed description of the optical data reduction and subsequent spectral analysis that allow us to derive the physical properties of this uniquely data-rich sample of galaxies.  The galaxy sample is divided into 38 pre-selected  L $\sim$ L*, z $\sim$ 0.2 ``target" galaxies and 29 ``bonus" galaxies that lie in close proximity to the QSO sightlines.  We report galaxy spectroscopic redshifts accurate to $\pm$ 30 km s$^{-1}$,  impact parameters, rest-frame colors, stellar masses, total star formation rates, and gas-phase interstellar medium oxygen abundances. When we compare the distribution of these galaxy characteristics to those of the general low-redshift population, we find good agreement. The L $\sim$ L* galaxies in this sample span a diverse range of color (1.0 $<$ $u-r$ $<$ 3.0), stellar mass (10$^{9.5}$ $<$ M/M$_{\odot}$ $<$ 10$^{11.5}$), and SFRs (0.01 $-$ 20 M$_{\odot}$ yr $^{-1}$). These optical data, along with the COS UV spectroscopy, comprise the backbone of our efforts to understand how halo gas properties may correlate with their host galaxy properties, and ultimately to uncover the processes that  drive gas outflow and/or are influenced by gas inflow.  
\end{abstract}


\section{Introduction}
\label{sec:intro}
  
  The gaseous halo is a key mediator between a galaxy and its
  intergalactic environment. Thus, establishing a basic set of
  observational facts about the physical state, metallicity, and
  kinematics of gas in the halos of galaxies is essential for
  understanding the nature of gas inflows and outflows thought to drive
  galaxy evolution. Nonetheless,  the large-scale gaseous  halos of galaxies have remained largely unexplored, due in part to observational challenges described below. 

  Theoretical investigations of galactic halos predict
  that a significant fraction of the medium should be diffuse and
  heated to a temperature characteristic of the virial mass of the
  underlying dark matter halo \citep{fraternali06}.  For galaxies like our own,
  this implies $T \gtrsim 10^6$\,K.  Such a hot, diffuse medium, even if
  metal-enriched, has a cooling time of order the Hubble time and
  therefore is unlikely to appreciably feed the galaxy's interstellar
  medium.   Inspired in part by the observations described below,
  modern treatments of galactic halos also envisage a cool phase $T
  \sim 10^4$\,K of gas, likely in pressure support with the hot phase
  \citep{mo96, maller04,keres05,dekel06}.  Indeed, this cool material is
  now predicted to fuel star-formation, the byproducts of which may potentially be 
  fed back to the IGM via galactic-scale outflows \citep{oppenheimer06}. Observational evidence of outflows is plenty \citep[e.g.][]{martin05,vcb05}, yet their significance to the course of galaxy evolution is undetermined. The outflowing gas may ultimately escape along with the metals generated in stars, or fall back down to the galaxy in a lather-rinse-repeat scenario. 
  
  Empirically, performing direct observations of gas in galactic halos has been a
  challenging exercise.  The medium is too diffuse and/or at a
  characteristic temperature that precludes detection in emission
  beyond the Galaxy and a handful of local systems
  \citep{bregman07}.  Regarding the Milky Way, 21cm surveys have revealed (for decades) a population of
  `high velocity clouds' (HVCs) at velocities inconsistent with rotation in the disk
  \citep[e.g.][]{munch61,wakker97}.  H$\alpha$ emission measures and carefully
  designed absorption-line experiments have now constrained these
  clouds to lie within the halo, at distances of $r \approx
  5-20$\,kpc \citep{weiner02,putman03,thom08,tripp11}.    These observations provide direct
   evidence of a cool medium within galactic halos.
  Furthermore,  a significant fraction of these HVCs exhibit \ovi\
  absorption implying the presence of a more highly ionized and most
  likely hotter medium \citep{fox04,sembach03}.  

  Beyond the Milky Way, one is essentially limited to exploring hot halo gas
  in absorption, i.e.\ by identifying bright background sources that
  coincidentally lie at close projected impact parameter to a
  foreground galaxy.  Because the overwhelming majority of diagnostic
  absorption-line transitions lie at rest-frame ultraviolet
  wavelengths, UV spectroscopy with spaceborne
  spectrometers is required to perform this type of  experiment at low redshifts.  The limited
  sensitivity of previous generations of instrumentation on the Hubble
  Space Telescope ({\it HST}) and the Far Ultraviolet Spectroscopic Explorer ({\it FUSE}) have yielded small
  samples of galaxies studied in this fashion. For instance, the pioneering work of  \cite{bowen95} describes a blind survey for Mg~II absorption in  17 background sightlines. While this study is unbiased by the previous knowledge of an identified MgII system, its focus is limited to a single ion.  
  
  Unlike the the work of \cite{bowen95},  the majority of absorption studies have not been conducted `blindly';  most absorbers were
  identified first in QSO spectra and a dedicated galaxy survey
  followed to associate a galaxy.  Such biases make it difficult to address questions about the origin of halo absorption and its dependence on galaxy properties. Thus, a clear understanding how the properties of halo gas relate to the properties of stellar populations has been elusive. Previous studies of Ly$\alpha$, C~IV, and Mg~II absorption lines indicate high covering fractions,  that gaseous halos have a large extent ($>$ 150 kpc), and that the properties of the gaseous halos are most likely governed by galaxy mass rather than a galaxy's star forming properties \citep{chen01, chen01b, chen10}.

  With the explicit goal of assessing the multiphase nature of halo
  gas in $L \approx L^*$, low-redshift galaxies, we have designed  and executed a large program
  with the {\emph{Cosmic Origins Spectrograph}}
  (COS; Froning \& Green 2009\nocite{froning09}) on {\it HST}.  Specifically, we are surveying the halo gas of 38 Sloan
  Digital Sky Survey (SDSS) galaxies (z = 0.15 $-$ 0.35) well inside
  their virial radii (with impact parameters $\rho$ $<$ 150 kpc). This
  COS-Halos survey obtains sensitive column density measurements of a
  comprehensive suite of multiphase ions in the spectra of 38
  z$<$ 1 QSOs lying behind  ``target" galaxies and a number of
  additional ``bonus" galaxies that happen to lie near the
  sightlines. In aggregate, these sightlines comprise a carefully-selecteed statistically sampled map of the physical state and metallicity of gaseous halos.  
  
  One key aspect of the COS-Halos survey is that it explores the
  variations of halo gas properties with galaxy properties. In order
  to obtain galaxy star formation rates (SFRs) and metallicities, the
  SDSS images of the galaxies are supplemented with high signal-to-noise,
  low-resolution optical spectra. Here, we describe the details of the optical
  observations and the spectral analyses that underscore the ``galaxy
  properties" side of  the COS-Halos survey as presented by Tumlinson
  et al. 2011. Recent work by \cite{thom11, meiring11, tumlinson11}
  showcase early results from this survey.   
    
  This paper proceeds as follows: Section 2 describes the
  low-resolution optical spectroscopy and data reduction; Section 3
  discusses the details of the spectral analysis that allows us to
  obtain SFRs and metallicities; and Section 4 presents the optical
  properties of these ``target" and ``bonus" galaxies. We refer the
  reader to Tumlinson et al. (in prep.) for a full presentation of the
  COS-Halos survey results, which includes a full analysis of gaseous
  halo properties compared to these optical galaxy properties.

\section{Foreground Galaxy Optical Spectroscopy}
 
 Tumlinson et al. (in preparation) provides the details of the
 QSO sightline selection for the COS large program, which we briefly
 summarize here. Relevant to this work, the targeted foreground
 galaxies in each sightline were selected to 
 (i) lie within 150 kpc projected separation from the sightlines, 
 (ii) have SDSS photometric
 redshifts ($z_{\rm phot}$ $-$1.5$\sigma$) that exceed 0.11 but are lower 
 ($z_{\rm phot}$ $+$ 1.5$\sigma$)  than the spectroscopic redshift of the QSO, and 
 (iii) have stellar masses between 10$^{10}$ $-$
 10$^{11}$ $M_{\odot}$ based on estimates from k-corrected SDSS
 $ugriz$ photometry. The redshift constraint (ii) was imposed to ensure that the OVI doublet would be redshifted into the COS bandpass, thereby providing a diagnostic of hot gas. Moreover, we emphasize that (i) and (iii) were based on SDSS photometric redshifts. Spectroscopic redshifts for all galaxies are included as part of the analysis presented here.  Approximately two-thirds of the foreground
 galaxies have blue colors ( $u - r$ $\lesssim$ 2.0) while the remaining third are red. 
 
\begin{figure}[ht]
\epsscale{1.0}
\plotone{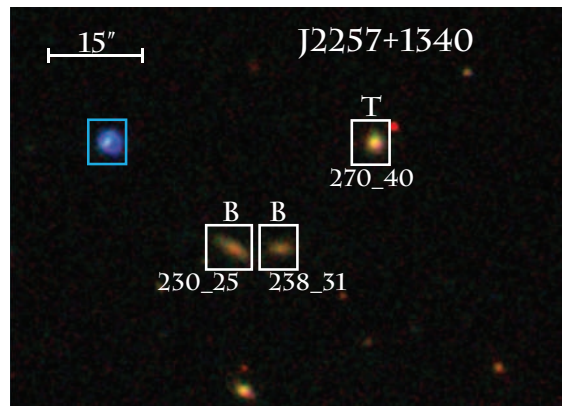}
\caption{ A three-color (g = blue; r = green; i = red) image of the field J2257+1340. The target (``T") and  bonus (``B") galaxies are marked and labeled by their identifiers.  \label{fig:finder}} 
\end{figure}

  In subsequent figures and tables, we identify individual galaxies by their 360-degree
 position angle (PA) from the QSO measured North to East, and their projected arcsecond
 separations ($\rho$\arcsec) in the form PA\_$\rho$\arcsec. Figure
 \ref{fig:finder} shows an example of the field surrounding the
 sightline at J2257+1340. In this case, the target galaxy, labeled with a ``T" is  270\_40. Two bonus galaxies, labeled ``B", 230\_25 and 238\_31 were also observed in this field. Typically, we selected a ``bonus" galaxy for follow-up spectroscopy based on (a) its proximity to the QSO being close enough to fit into the Keck LRIS longslit with the target galaxy and/or (b) a photometric redshift that matched the target galaxy criteria, (ii; above), or that of an additional absorber we already detected in the  QSO sightline. While the target galaxies represent a carefully selected blind sample, the bonus galaxies are a heterogenous, absorption-biased sample. We analyze the properties of the target and bonus galaxies separately in this work.

In total, we obtained longslit, optical spectra for each of the 38
 target galaxies and 29 bonus galaxies 
 over the course of six different observing runs at two telescopes, Keck I and  Magellan II
 Clay. On the Keck I telescope, we used the Low-Resolution Imaging
 Spectrometer (LRIS), while on the Clay telescope, we used the
 moderate-resolution Magellan Echellete (MagE) spectrometer. Both
 instruments provide full coverage of the optical spectrum between
 approximately 3100 \AA~ and 9000 \AA. Table \ref{tab:obs} summarizes
 the observing runs. Below, we provide details about the observations
 made with each instrument.

 \begin{table*}[hpt]\centering \scriptsize
\begin{tabular*}{0.7\textwidth}{@{\extracolsep{\fill}}lccccc}
\hline
Run & Instrument & Grating(s) &$\lambda_{cen}$ Blue, Red [\AA]& Slit & N$_{gal}$ \\
\hline
(1) & (2) & (3) & (4)  \\
\hline
\hline
October 2008 & LRIS &600/4000, 600/7500 &4323, 6905 & 1.0\arcsec&7\\
March 2009 & LRIS & 600/4000, 600/7500  &4323, 6905 &1.0\arcsec &7\\
March 2010 & LRIS & 600/4000, 600/7500  &4323, 7220 & 1.0\arcsec&25\\
April 2010 & LRIS & 600/4000, 600/7500  &4323, 7056 & 1.0\arcsec&21\\
March 2011 & MagE & 175 gr/mm& 6200 & 0.7\arcsec &6 \\
May 2011 & LRIS &  600/4000, 600/7500 &4353, 7120 & 1.0\arcsec&1\\

\hline

  \hline
\end{tabular*}
\caption[Summary of Observing Runs]{ Summary of observing runs. For
  each grating used, we give the central wavelength in column 3. For
  LRISb, the spectral coverage is fairly constant at $\sim$ 3100 $-$
  5600 \AA. On LRISr, the wavelength coverage begins between 5600 $-$
  5800 \AA~ while the maximum wavelength ranged from 8200 $-$ 8800
  \AA. For MageE, the spectral coverage is continuous for 3200 $-$
  10000 \AA.  \label{tab:obs}} 
\end{table*}

  \subsection{Keck LRIS Data}
   Over the course of several observing runs (October 2008,
  March 2010, April 2010, and May 2011) using the Keck I 10-m telescope
  Low-Resolution Imaging Spectrometer (LRIS; Oke et
  al. 1995\nocite{oke95}), we obtained spectra of 61 galaxies (35 target galaxies and 26 bonus galaxies) along 35 QSO sightlines.  For these LRIS data, we use a 1 \arcsec~slit,  the D560 dichroic with the 600/4000 l/mm grism (blue side)
  and 600/7500 l/mm grating (red side). Binning the data 2 $\times$ 2
  on the blue side and 1 $\times$ 2 (spatial $\times$ spectral) on the red side gives dispersions
  of 1.2 and 2.3 \AA~per pixel, respectively.  Exposure times varied according to
  galaxy brightness and sky conditions, but were generally  sufficient
  for obtaining signal-to-noise ratios of at least 3 per pixel for
  strong nebular emission lines in the galaxy spectra.  Table \ref{tab:keck_obs} provides some of the observational parameters for the LRIS spectra for each galaxy, including the date observed, exposure times, apparent magnitudes, and flux correction factors (discussed below).
  
  The longslit was typically oriented at a PA to include our target
  galaxy and either the background quasar or an additional galaxy at close impact
  parameter to the sightline.  We use the LRIS Cassegrain Atmospheric Dispersion Compensator (ADC) to minimize
  light-loss from atmospheric dispersion.  Table~\ref{tab:keck_obs}
  summarizes the targets and individual exposures.
  
  In addition to the science observations, we acquired a series of
  calibration images on each night.
  Spectral flats on the blue side consist of slitless pixel-flats
  taken during twilight, which represent the intrinsic pixel-to-pixel
  response variations of the CCD, and twilight flats with
  1\arcsec~slit, which represent the larger scale illumination
  variations due to non-uniformities in the width of the slit and
  vignetting. On the blue side, we use the twilight sky for spectral
  flats because the dome lamps emit too little UV light. On the red
  side, we use the dome flats for both the pixel-to-pixel calibration
  and the illumination correction since these lamps reduce the level
  of scattered light. In addition to these spectral flats, we also
  observed a set of arc lamps at the beginning and end of every night
  for wavelength calibration, and at least one spectrophotometric
  standard star per night for flux calibration.  

  The two-dimensional spectral images were reduced with the
  LowRedux\footnote{http://www.ucolick.org/~xavier/LowRedux/index.html}
  pipeline, developed by J. Hennawi, D. Schlegel, S. Burles, and JXP.
  The pipeline bias-subtracts each exposure, generates a flat-field
  frame from the calibration images, and generates a two-dimensional
  wavelength image (pixel-by-pixel) from the arc lamp exposures.  The
  code automatically identifies sources in the slit, masks these
  objects, and calculates a global estimate of the sky background from
  the remaining pixels.  In the majority of cases, this sky solution
  is refined to be localized to each source during extraction.  In several
  cases, however, we found better results from the global solution
  alone (especially for pairs of objects in close proximity).  

  The final 1D spectra were optimally extracted from the
  two-dimensional images, and multiple exposures of a given target
  were co-added, weighting by the inverse variance.  The wavelength solution of
  these spectra were corrected for instrument flexure through a
  comparison of the sky spectrum with an archived solution.  The
  wavelengths were then shifted to a vacuum and heliocentric reference
  frame.   

  Spectral fluxing was performed in several steps \citep[see
  also][]{dasilva11}.   An initial estimate for the flux was made
  using a sensitivity function generated from our observations of
  spectrophotometric standard stars.  We expect that this provides a
  good estimate for the relative flux within each camera, but it
  does not properly account for slit-losses.  
  
  To bring the spectra to an absolute flux scale, we convolved the LRISb spectrum with the SDSS $g$-band filter response curve and scaled the
  flux of the blue spectrum to match the reported SDSS-DR7 petrosian
  magnitude.  Similarly, we matched the
  $i$-band magnitude for our LRISr spectra. The median values of these
  two flux factors are 1.94 (blue) and 1.77 (red), corresponding to
  median slit light losses of 48\% and 43\%. For a small subset of our
  sample (seven galaxies), these factors are $> 5$. Moreover, the
  scale factors are occasionally discrepant between the blue and the
  red sides by more than a factor of 2 (six galaxies, four of which
  fall into the previous subset). The large and/or discrepant flux
  scale factors are due to a diverse set of factors:  
  spatially extended galaxies (i.e.\
  systems where slit-loss is extreme), very faint
  galaxies with high photometric errors, at least one case of probable
  poor slit alignment, galaxies with close neighbors in projection,
  and cases of poor seeing $>$ 2\arcsec.  The SDSS magnitudes and
  corresponding flux scale factors are listed in Table
  \ref{tab:keck_obs}.   
  
  Lastly, we applied a correction for Galactic extinction, assuming the $E(B-V)$ value from
  \cite{schlegel98} and a Galactic extinction law \citep{ccm}.
  The resultant spectra of the target  and bonus galaxies are presented in
  Figures~\ref{fig:keck_target_spec} and ~\ref{fig:keck_bonus_spec}.  

 \subsection{Magellan MagE Data}

In March 2011, three remaining target galaxies and several additional bonus galaxies were observed with the Magellan Echellette (MagE) spectrograph on the Magellan Clay telescope at Las Campanas Observatory \citep{marshall08}. MagE provides moderate resolution spectral coverage between 3200 \AA~and 10000 \AA. These data were acquired with a 0.7\arcsec~ slit and binned 1 $\times$ 1, giving dispersions of 0.3 \AA~ per pixel at [OII] $\lambda\lambda$ 3727, and 0.5 \AA~ per pixel at H$\alpha$.  The average emission-line FWHM is 55 km s$^{-1}$ for these data.  Exposure times varied between 600 and 1200 seconds, depending on the target galaxy apparent magnitude. Table \ref{tab:mageobs} provides some of the observational parameters for the MagE spectra for each galaxy, including the date observed, exposure times, apparent magnitudes, and flux correction factors.

The MagE spectra were reduced using the MASE
pipeline\footnote{http://web.mit.edu/jjb/www/MASE.html  \citep{bochanski09}} developed by
Bochanski, Simcoe, and Hennawi, which is an adaptation of the
MIKE/HIRES echelle extraction codes. 1D spectra
are optimally extracted from the 2D reduced images. Several
spectrophotometric standard stars taken at a variety of airmasses were
used to initially flux-calibrate the data. As with the LRIS data, we
account for slit-losses by scaling the spectra to match SDSS
photometry. Because there is no dichroic in the MagE data, we correct
these spectra with the SDSS $r$-band photometry. Once we scale the spectra by the
appropriate flux factor, we confirm that the resultant colors in the
SDSS bands match to within 0.2 magnitudes and conclude that the $r$-band normalization is approximately correct for $u$ and $g$ bands as well.

\section{Analysis}
\label{sec:anal}

\subsection{Redshift Determination}
\label{sec:redshift}

\begin{figure*}[hpt]
\centering
\subfigure{
\includegraphics[height=0.48\linewidth,angle=90]{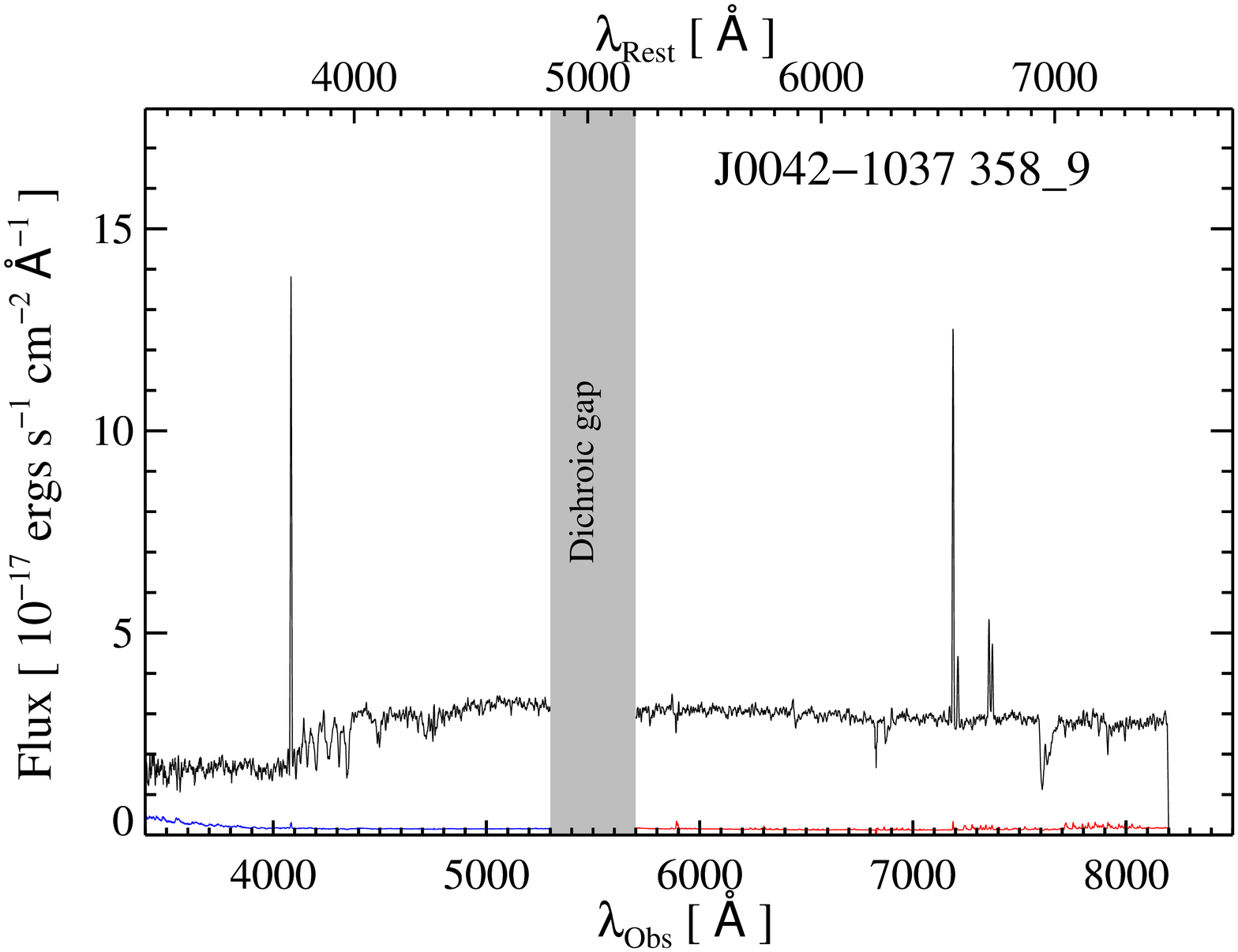}}
\subfigure{
\includegraphics[height=0.48\linewidth,angle=90]{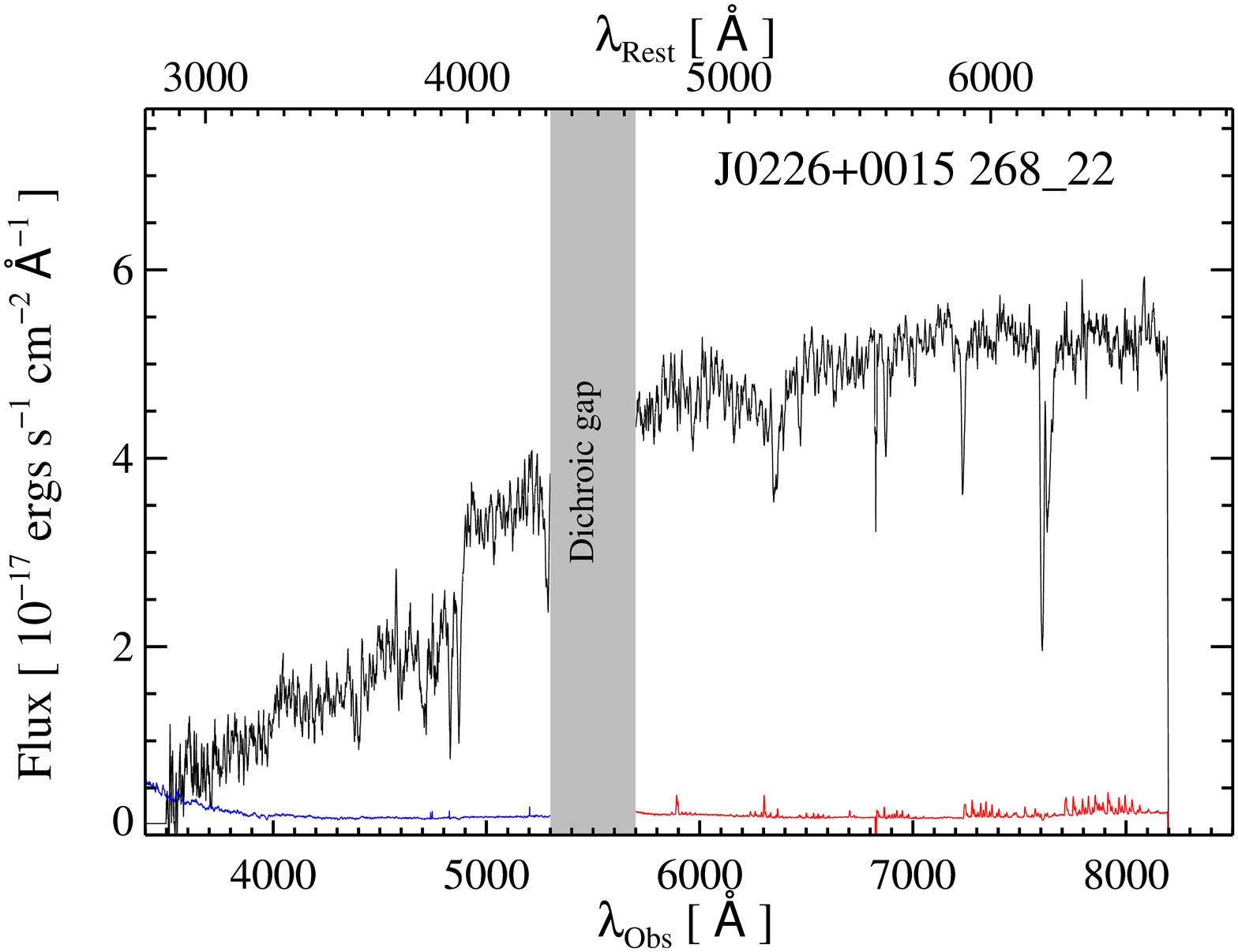}}
\subfigure{
\includegraphics[height=0.48\linewidth,angle=90]{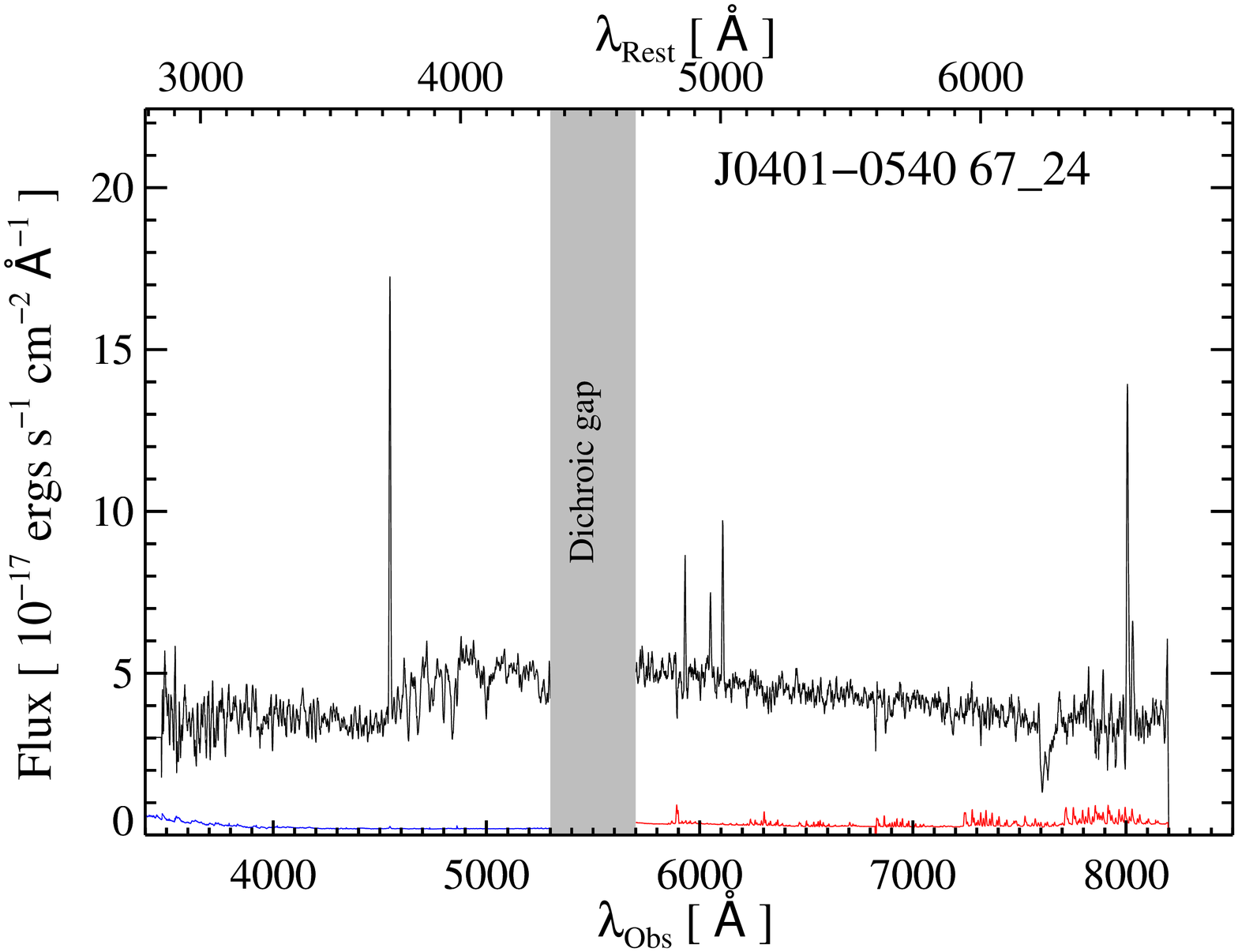}}
\subfigure{
\includegraphics[height=0.48\linewidth,angle=90]{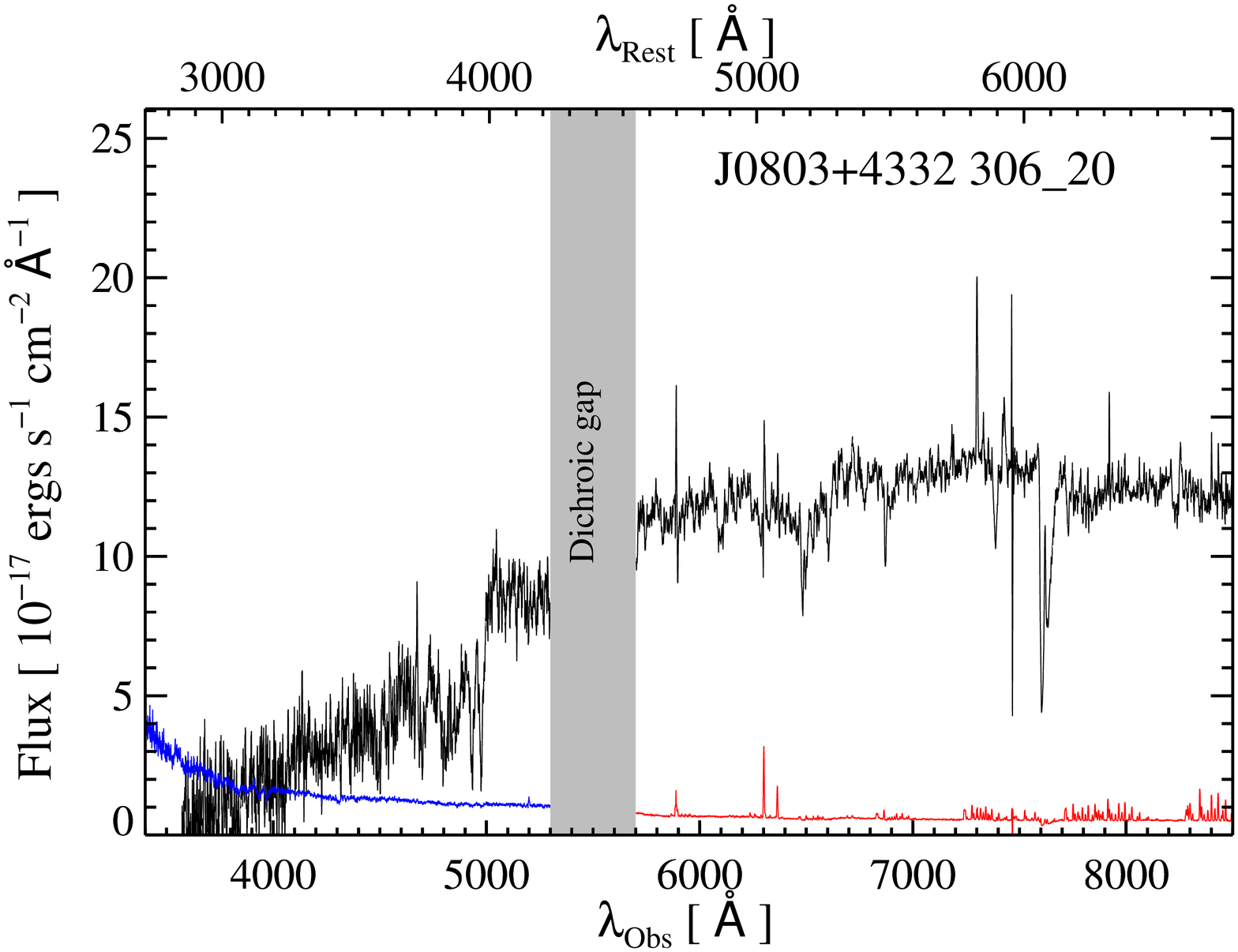}}
\subfigure{
\includegraphics[height=0.48\linewidth,angle=90]{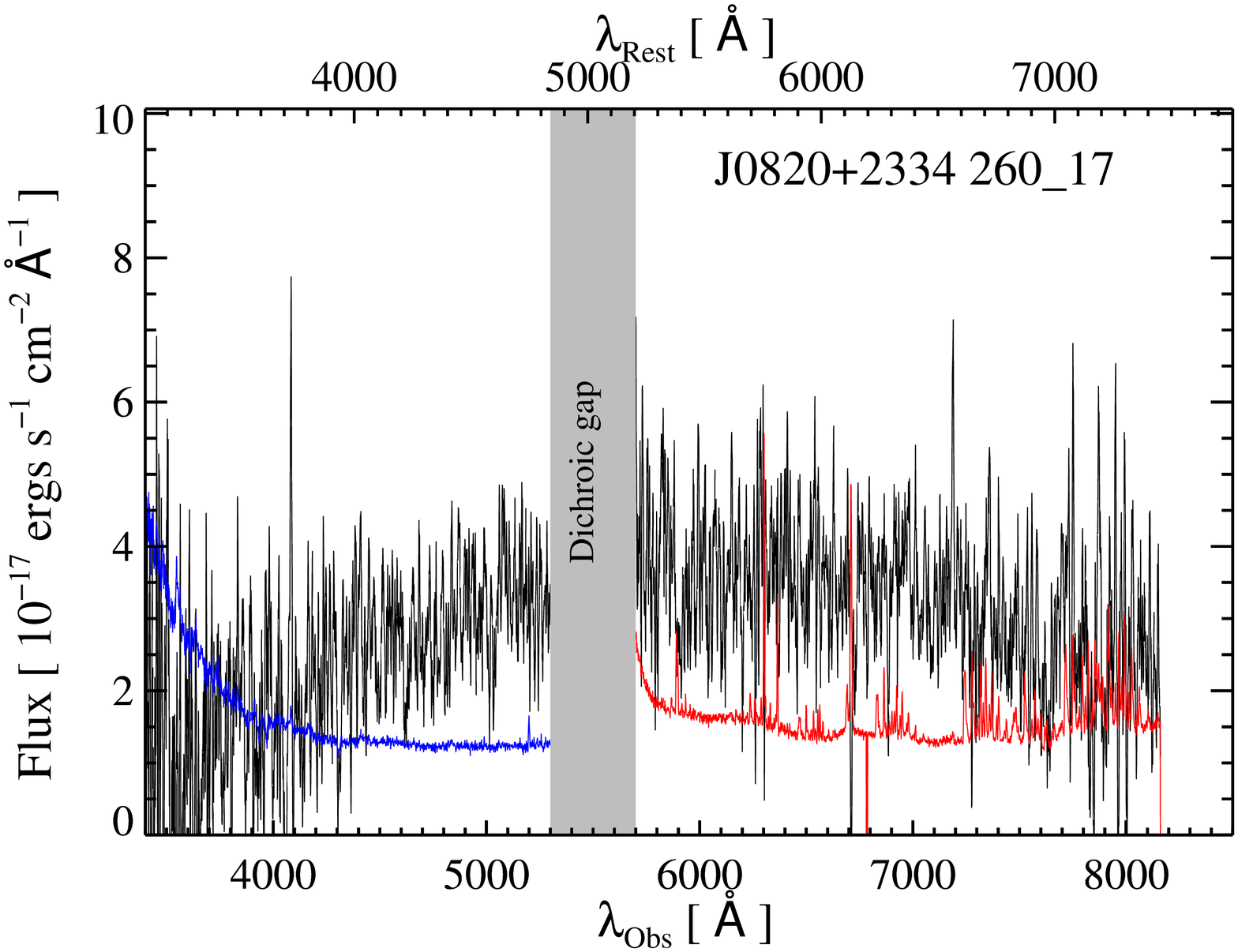}}
\subfigure{
\includegraphics[height=0.48\linewidth,angle=90]{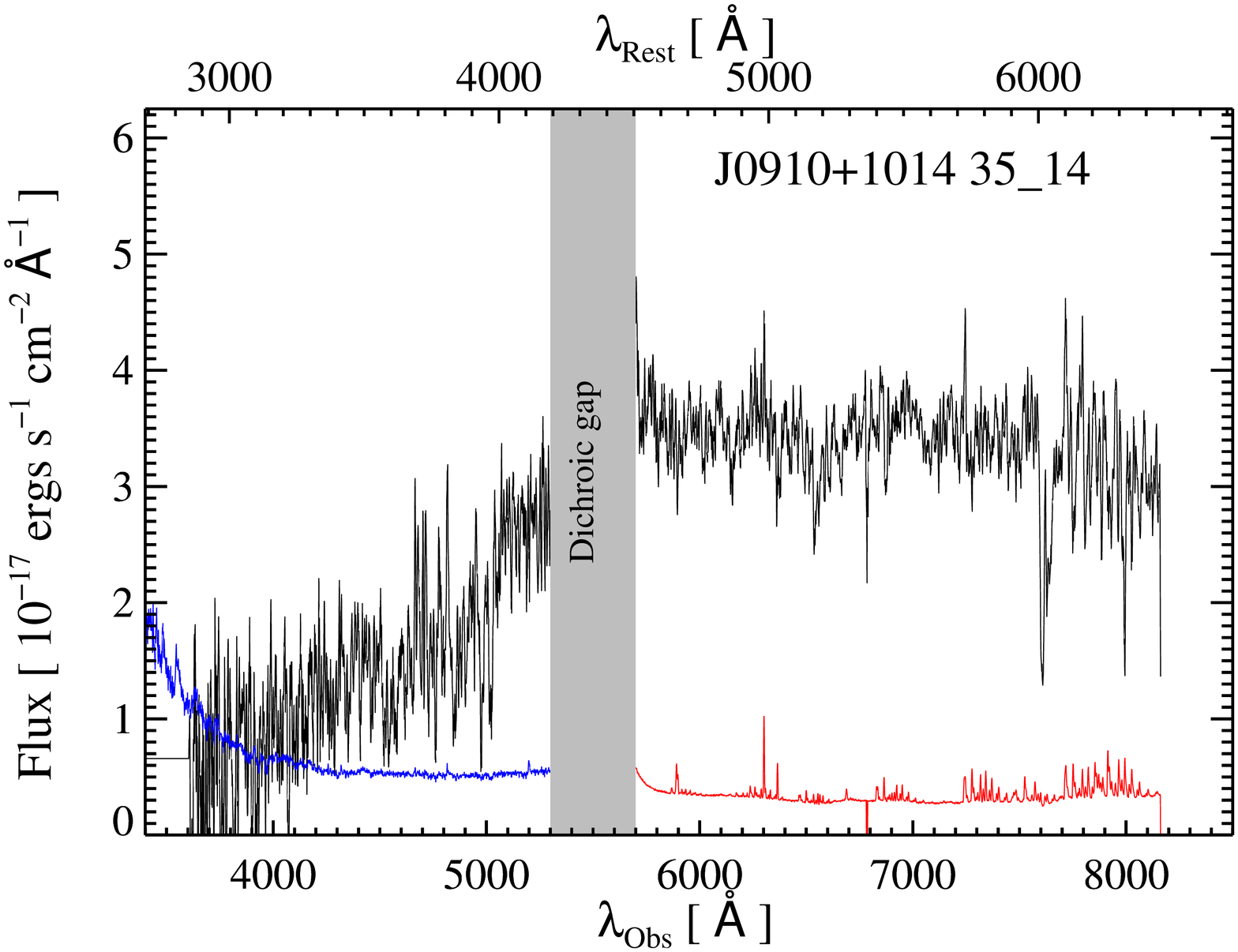}}
\caption{The 1D reduced, flux-calibrated spectra for selected target galaxies. All spectra will be available in the online, published version.We represent the dichroic with a shaded area  near the observed wavelength 5000 \AA.  \label{fig:keck_target_spec}}
\end{figure*}

\begin{figure*}[hpt]
\centering
\subfigure{
\includegraphics[height=0.48\linewidth,angle=90]{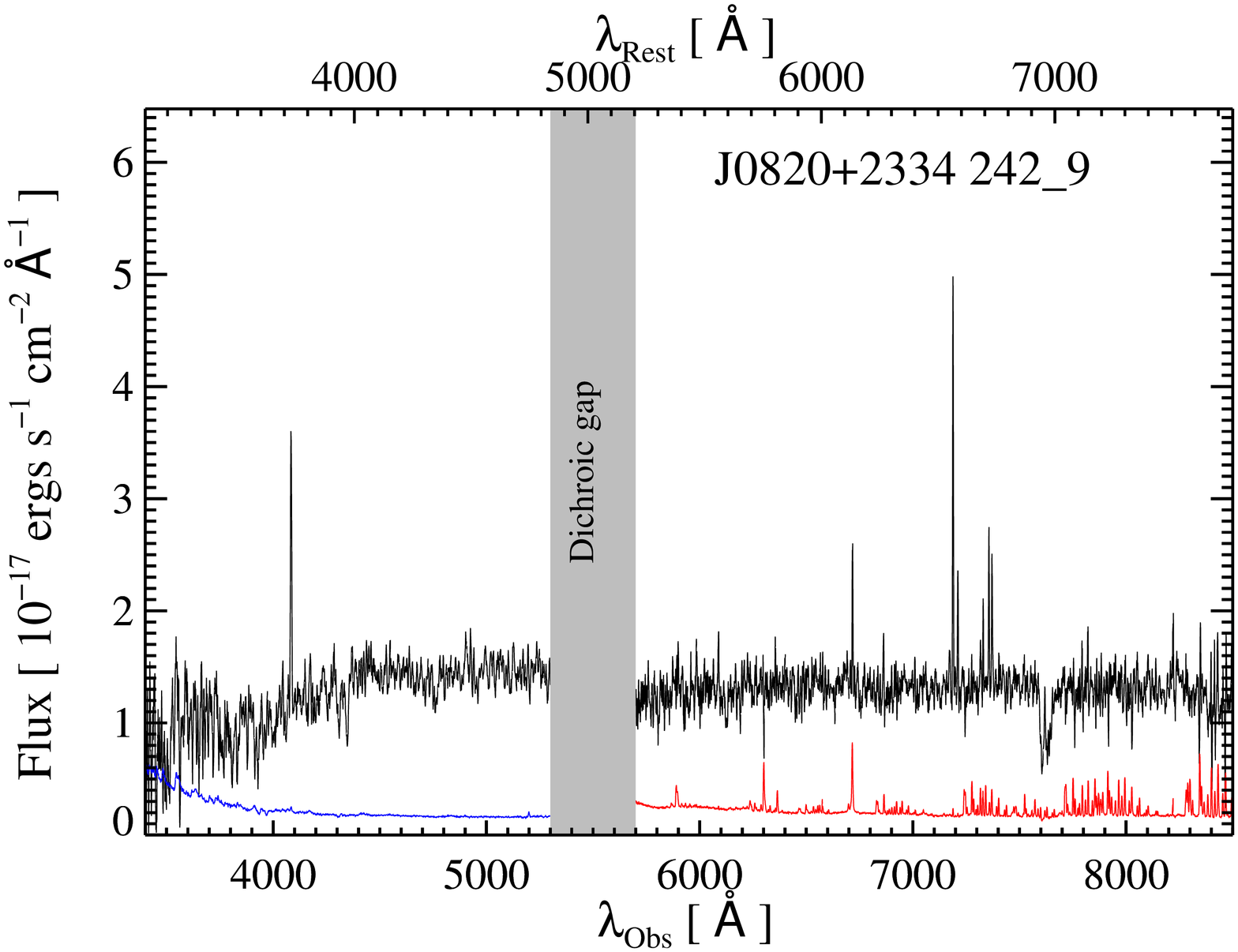}}
\subfigure{
\includegraphics[height=0.48\linewidth,angle=90]{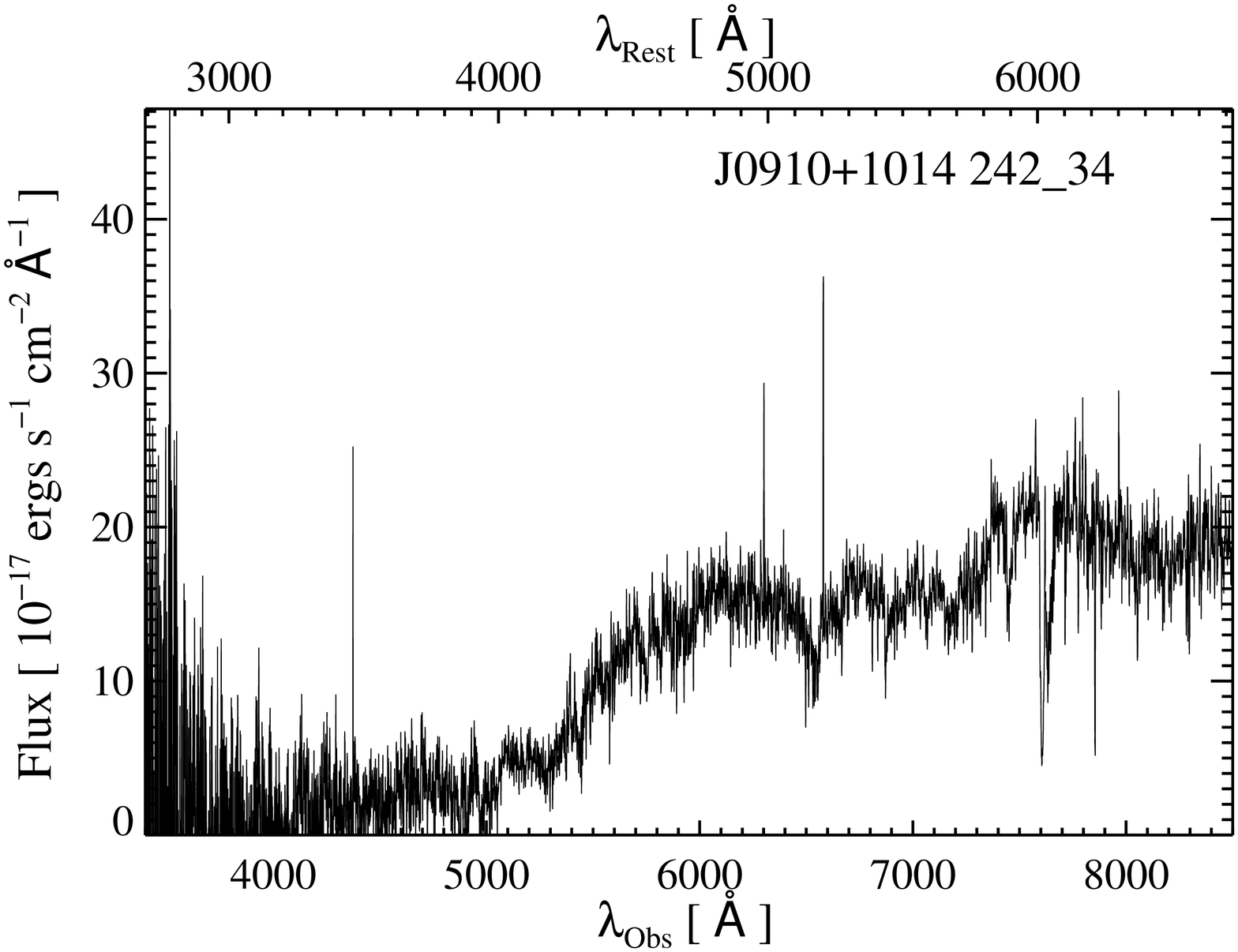}}
\subfigure{
\includegraphics[height=0.48\linewidth,angle=90]{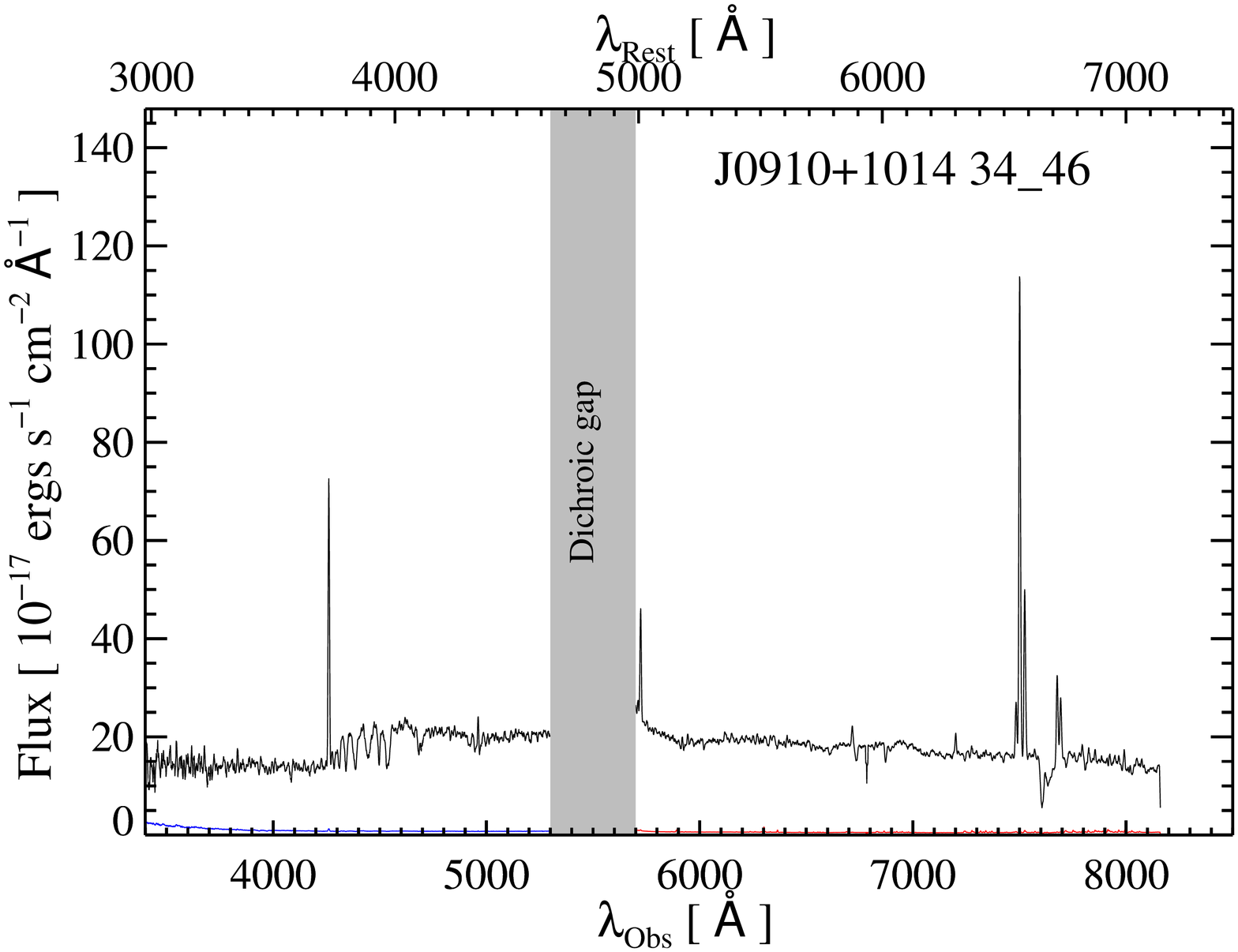}}
\subfigure{
\includegraphics[height=0.48\linewidth,angle=90]{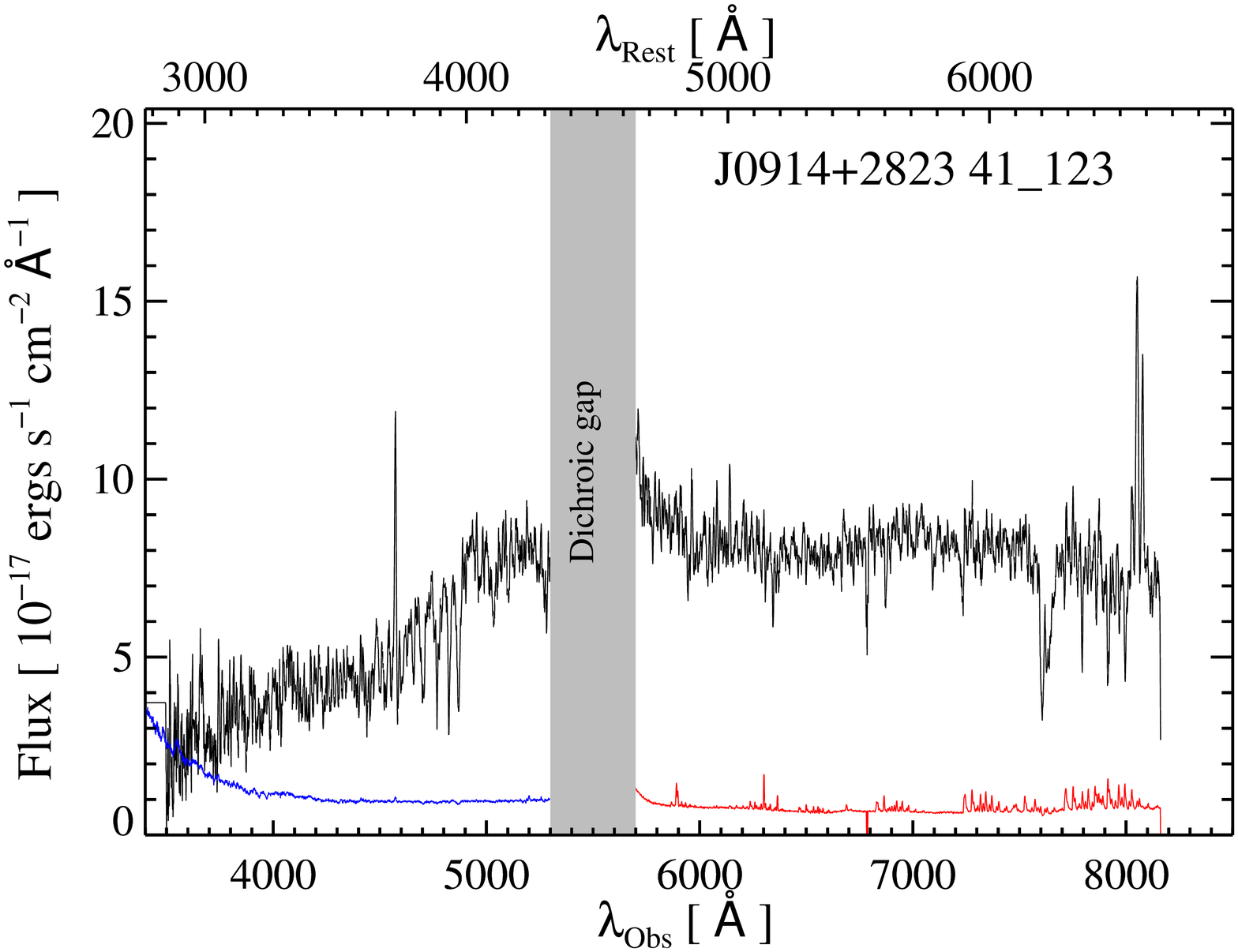}}
\subfigure{
\includegraphics[height=0.48\linewidth,angle=90]{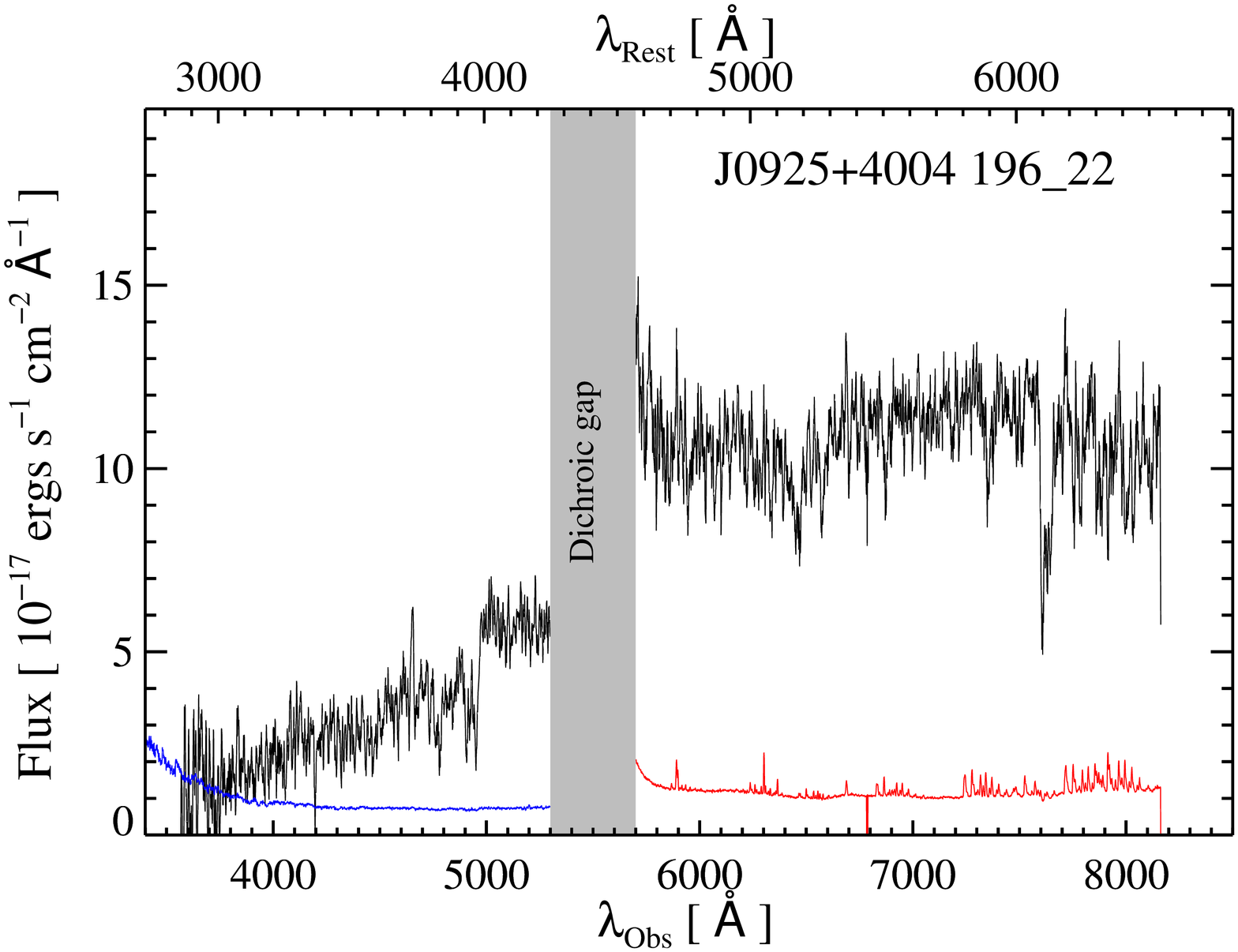}}
\subfigure{
\includegraphics[height=0.48\linewidth,angle=90]{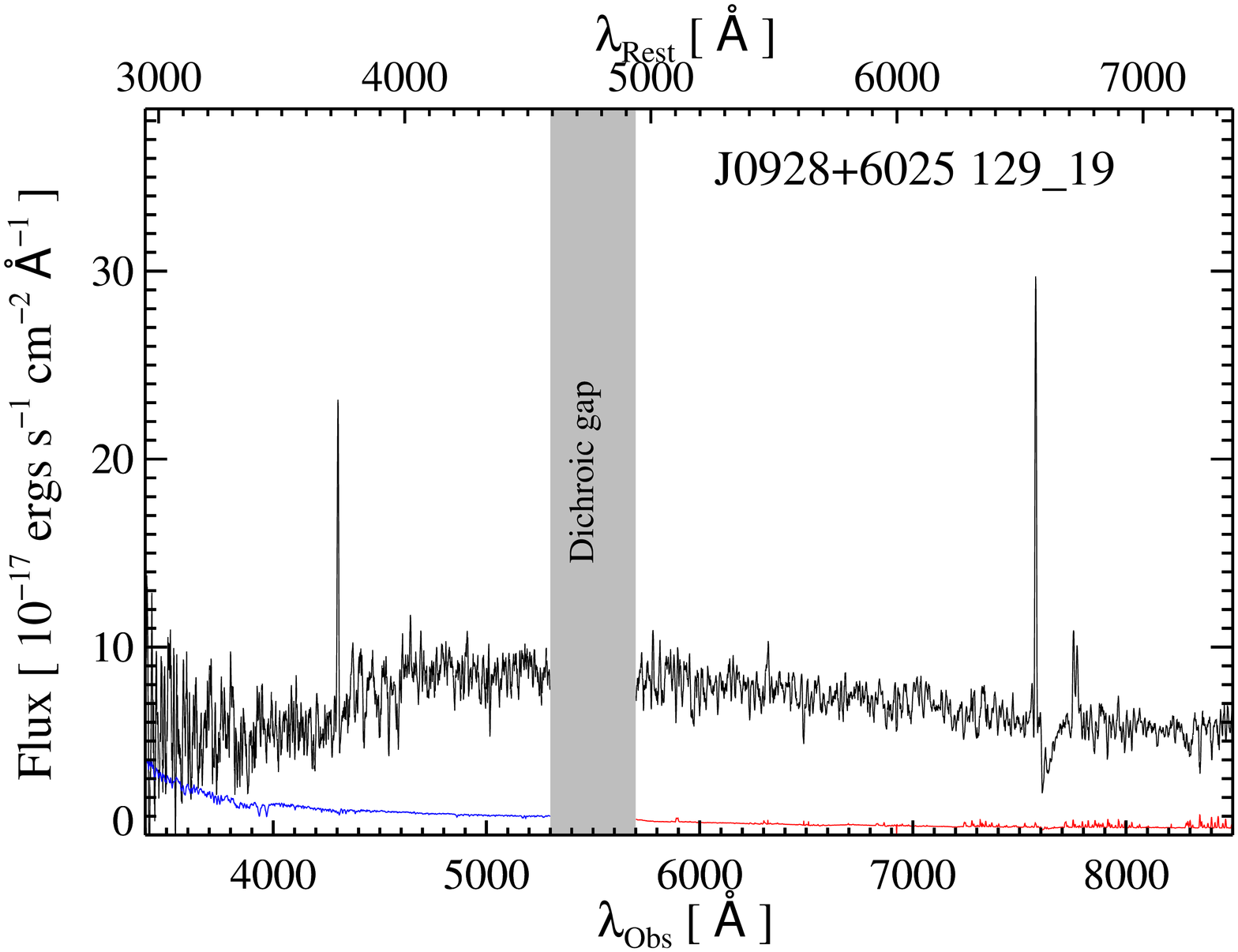}}
\caption{The 1D reduced, flux-calibrated spectra for selected bonus galaxies. All spectra will be available in the online, published version. We represent the dichroic with a shaded area  near the observed wavelength 5000 \AA. \label{fig:keck_bonus_spec} }
\end{figure*}

It is of primary scientific interest to our program, which examines the gas
in galactic halos, to establish a precise redshift for each target
galaxy.  One can then search the spectra of background quasars for any
coincident absorption.  All of our target galaxies exhibit significant
absorption lines (e.g.\ Ca H+K) and/or emission lines (e.g. H$\beta$)
which provide a precise redshift measurement for stars and the ISM
(Figure~\ref{fig:keck_target_spec} and \ref{fig:keck_bonus_spec}).  Instead of analyzing these
spectral features individually, we employed a modified version of the
SDSS algorithm {\it {zfind}} that is bundled within the IDLUTILS
package.{\footnote{http://spectro.princeton.edu/idlspec2d\_install.html}}
In brief, the code models the input LRIS spectrum using a set of
archived Principle Components Analysis (PCA) eignevectors derived from galaxies observed in the SDSS
survey.  We also include a model of the instrumental spectral resolution, allowing and solving for internal dispersion within the
galaxy.  The code calculates the $\chi^2$ in steps of redshift space,
reports the minimum value, and provides an estimate of the redshift
uncertainty.  

For the Keck/LRIS observations, we performed this analysis on the red
and blue sides of the spectra separately.   In two cases of very faint galaxies at low redshift, only the H$\alpha$ emission line is present in the LRISr spectrum and $zfind$ fails on the red side. In these two cases we manually entered the red-side redshift to be equal to that of the blue side, which was based on [OII], H$\beta$ and [OIII]. As the statistical uncertainties reported by {\it {zfind}} are generally $<$~5 km s$^{-1}$, the precision of our redshift
measurements is limited by systematic uncertainty. The RMS of the wavelength solution and the flexure correction are the two primary sources of error. The two independent redshift determinations on the red and blue sides offer some insight into its magnitude. 

\begin{figure*}[ht]
\epsscale{0.85}
\plotone{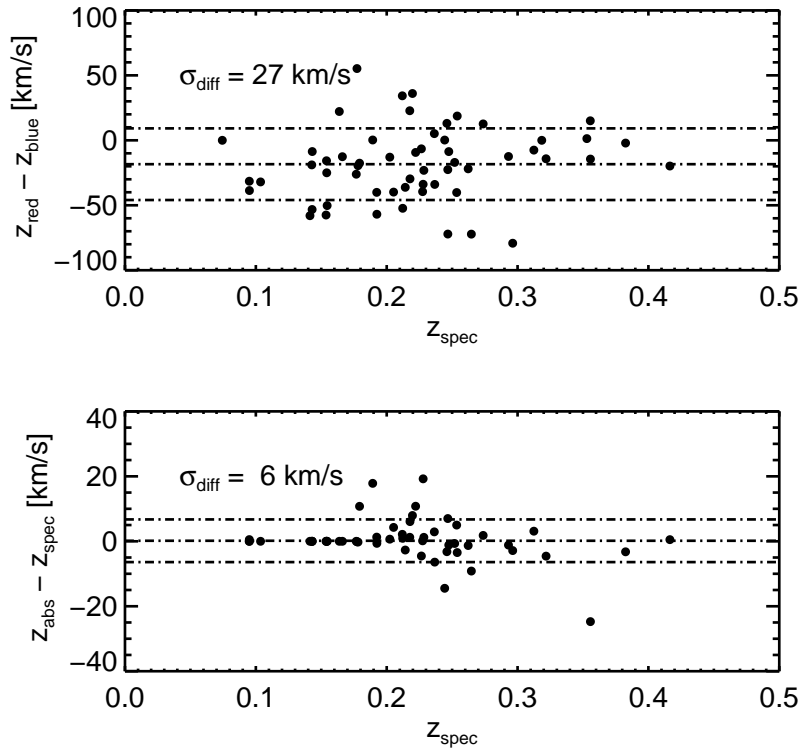}
\caption{ Top: velocity difference between blue and
  red-side spectral redshifts. Bottom: Velocity difference between redshifts measured using
  all spectral features, and redshifts measured after masking out
  emission lines. The red-side spectral redshift using
  the full spectrum is on the x-axes of both plots. Dashed-dotted lines
  mark the standard deviation (top and bottom lines) and mean (middle
  line) of the distribution of these differences.  \label{fig:zcomp}}
\end{figure*}

The top panel of Figure \ref{fig:zcomp} compares the resultant LRISb
and LRISr {\it{zfind}} redshifts in velocity space.  This plot shows that the redshifts
derived from the blue camera tend to be systematically higher than those derived from the red
camera by nearly 10 km s$^{-1}$. Such an offset may result from: 
(1) The difference in the instrumental flexure correction for the blue
and red cameras, and  
(2) the use of the unresolved [OII] $\lambda$$\lambda$ 3727
doublet in the redshift determination of the blue camera.  The [OII] emission line is
often the strongest spectral feature in that spectrum. Since the
flexure correction is done with night-sky lines, the blue side is
subject to higher uncertainty owing to there being far fewer sky lines
below 5000 \AA. Taking into account both of these effects, we are
inclined to trust the redshifts from the red side over the blue. The
resultant redshifts from LRISr are listed in the third column of
Table~\ref{tab:galprops}. We estimate the overall uncertainty of the
redshifts given in the table by the standard deviation of the redshift
differences between the red and blue sides seen in Figure
\ref{fig:zcomp}. Thus, we adopt a conservative  30 km s$^{-1}$
systematic uncertainty in our final redshift measurements. This uncertainty is
primarily due  to wavelength calibration error (a combination of RMS
in the arc-line analysis and instrument flexure).  

To determine the precise redshifts of the galaxies observed with
Magellan MagE, we use the same SDSS {\it{zfind}} algorithm for the
entire spectrum. Thus, the redshifts we report in
Table~\ref{tab:galprops} for the MagE galaxies were made using the
entire spectrum. Because the spectral resolution of MagE is higher
than that of LRIS (the [OII] $\lambda$$\lambda$ 3727 doublet is
resolved in these spectra), the systematic uncertainty in the MagE
redshifts is lower than that of the LRIS redshifts. We estimate the
uncertainty of these redshifts to be 5 km s$^{-1}$, based on the RMS
of the wavelength calibration.  
 
We made one final check on our redshift estimates following
\nocite{rubin11} Rubin et al. 2011, who note modest, but
non-negligible, offsets between emission lines and stellar absorption
features in spectra of $z \sim 0.5$ galaxies.  
The bottom panel of Figure~\ref{fig:zcomp} compares the redshift
measurements from {\it{zfind}} made using the entire LRISr spectrum
(z$_{spec}$) against a similar analysis but with galaxy emission-line
features masked out (z$_{abs}$). 
There is generally good agreement between z$_{spec}$ and
z$_{abs}$, with no significant systematic offset between the two
measurements. Furthermore, the standard deviation of this distribution
(6 km s$^{-1}$) falls well-within the adopted 30 km s$^{-1}$
systematic redshift uncertainties.   Therefore, we do not consider
this a significant concern for our analysis.

\begin{figure*}[h]
\epsscale{0.75}
\plotone{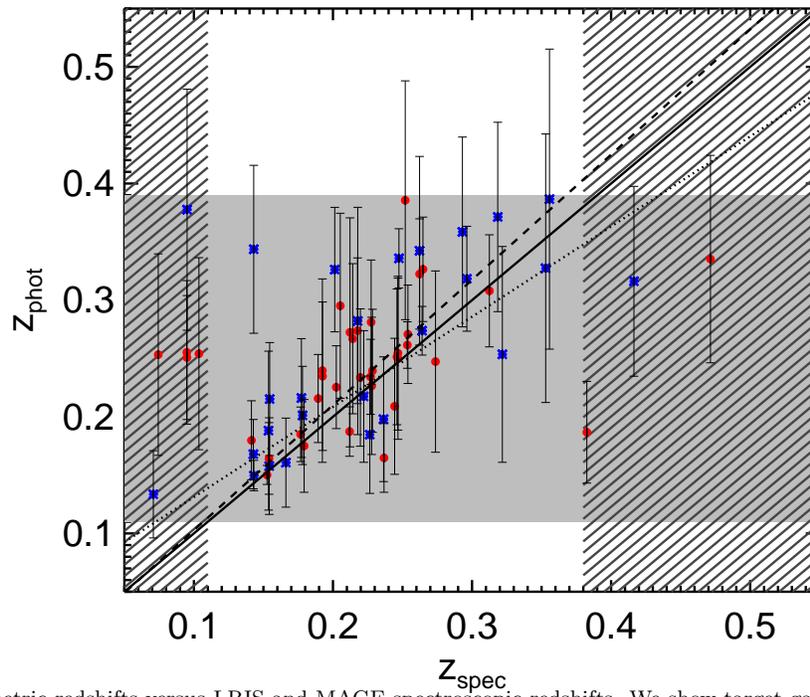}
\caption{ SDSS photometric redshifts versus LRIS and MAGE spectroscopic redshifts. We show target galaxies as red filled circles and bonus galaxies as blue asterisks. The shaded area of the plot highlights the original sample selection criteria described in Tumlinson et al. 2011, while the hashed area of the plot highlights those objects with spectroscopic redshifts that ultimately fall outside the selected redshift range. The one-to-one line is shown as a  solid line. The dotted line represents a fit to all the data, while the dashed line represents a fit to the data in the range 0.11 $<$ z$_{\rm spec}$ $<$ 0.38.  \label{fig:zphot}}
\end{figure*}

In Figure \ref{fig:zphot} we plot the galaxy SDSS photometric redshifts (z$_{\rm phot}$) versus their spectroscopic redshifts (z$_{\rm spec}$). For reference, the values of  z$_{\rm phot}$ that we use here are part of the ``Photoz2" online catalog, and called ``photozd1." These redshifts and their errors are calculated using SDSS galaxy magnitudes and a Neural Network method \citep{oyaizu08}. The gray shaded region of this plot highlights the original COS-Halos survey redshift selection criterion, while the hashed area is shown to mark the region of spectroscopic redshift parameter space that ultimately falls outside of the pre-selected range. In order the assess the accuracy of the z$_{\rm phot}$, we calculate  $\chi^{2}$ assuming that the relation between z$_{\rm phot}$ and z$_{\rm spec}$ should be a one-to-one linear correlation. We find that  $\chi^{2}$ is quite large in this case (n$_{\rm dof}$ = 64; excludes the two galaxies in the sample not identified by SDSS), with a value of $\sim$115, and has an associated probability of $\sim$0.1\%.  If we exclude those points that fall within the hashed area of this plot, $\chi^{2}$ is lowered considerably to $\sim$55 (n$_{\rm dof}$ = 53), corresponding to a probability of $\sim$40\%. Thus, the large $\chi^{2}$ for the full sample is driven by a handful of ``catastrophic failures." On this plot, we also show error-weighted linear-least-squares fits to the data using the full sample (dotted line) and the constrained sample in which 0.11 $<$ z$_{\rm spec}$ $<$ 0.38 (dashed line). The fits to the two samples are: z$_{\rm fit}$ = (0.05 $\pm$ 0.01) + (0.77 $\pm$ 0.07)$\times$z$_{\rm spec}$, with a $\chi^{2}$ of 99.25 (P = 0.24\%), and z$_{\rm fit}$ (0.11 $<$ z$_{\rm spec}$ $<$ 0.38) = ($-$0.005 $\pm$ 0.02) + (1.07 $\pm$ 0.09)$\times$z$_{\rm spec}$ , with a $\chi^{2}$ of 49.6 (P = 56.8\%). 

\subsection{K-corrections: Stellar Masses and Absolute Magnitudes}
\label{sec:mass}
  
  To obtain an estimate of the current stellar mass and absolute magnitudes of each galaxy, we
  used version 4\_2 of Michael Blanton's IDL package 
  $kcorrect$\footnote{found at http://howdy.physics.nyu.edu/index.php/Kcorrect}
  \citep{blanton07}, the SDSS DR7 galactic reddening corrected,
  asinh $ugriz$ magnitudes, and the $zfind$ spectral redshifts. Specifically, we use the 
  routine $sdss$\_$kcorrect$ to obtain a suite of distance-dependent galaxy properties for every galaxy in our sample. Two
  exceptions are the bonus galaxy J1009+0713: 86\_4 and 
  target galaxy J1157-0022: 230\_7
  which lie at very close impact parameters to the QSO and are not
  identified as separate galaxies in the SDSS catalogs. The stellar
  masses and absolute magnitudes that are output by $sdss$\_$kcorrect$
  contain a factor of $5 \log h$, with the unitless $h = 1$. The
  masses and absolute magnitudes we adopt throughout this work have
  been corrected assuming the 5-year WMAP cosmology with a $h=0.72$
  \citep{dunkley09}.  
  
  For red galaxies that are not detected by SDSS in the $u$-band, we put
  realistic flux-based limits on the absolute $u$-band magnitude rather
  than employ the output k-corrected asinh absolute magnitude (which
  may even be based on a negative flux). The procedure for determining
  these limits is as follows: (1) we use the routine $k\_lups2maggies$
  to convert SDSS luptitudes (asinh magnitudes) to maggies (a
  flux-like quantity), (2) we then input the maggies and their inverse
  variance into $sdss\_kcorrect$ to obtain absolute magnitudes (M) and
  their inverse variances (M$_{\rm ivar}$), (3) we determine the
  corresponding absolute maggie (F$_{\rm maggie}$) and its inverse
  variance (F$_{\rm ivar}$), taking F$_{maggie}$ = 10$^{-0.4\rm M}$
  and F$_{\rm ivar}$ = M$_{\rm ivar}$ / (0.4 $\times$ ln10 $\times$
  F$_{\rm maggie}$)$^{2}$, (4) We determine an absolute magnitude
  limit (M$_{\rm lim}$) in the $u$-band for each individual galaxy
  such that M$_{\rm lim}$ = -2.5 $\times$ log$_{10}$ (2 $\times$
  (1/F$_{\rm ivar}$)$^{1/2}$).  If a detection of a galaxy in the
  $u$-band 
  is less than 3$\sigma$, we adopt a 2$\sigma$ 
  magnitude limit and report a lower limit to the $u-r$ color. 

  The middle-left panel of Figure \ref{fig:hist} shows the
  distribution in log space of stellar masses for the full
  sample of foreground galaxy masses (light shade) and for only the
  target galaxies (dark shade). These stellar masses are also listed
  in Table \ref{tab:galprops}. The median log $M_{*}$ is 10.31, in a
  distribution that ranges from $\sim$ 8.8 $-$ 11.3. The mass
  distribution of SDSS galaxies (based on photometric redshifts and
  stellar absorption line indices) shows a bimodal distribution with a
  break near 10.4, above which there is an increasing fraction of
  older-population elliptical galaxies \citep{kauffmann03}. 
  Our sample brackets this break but contains very few systems with
  $M_* \ll 10^{10} \mmsun$.  This bias results from our sample selection
  criteria that were meant to isolate L* galaxies.  We selected galaxies from SDSS that had apparent  magnitudes sufficiently bright to provide a precise photometric
  redshift $z_{\rm phot}$ and also yielded $z_{\rm phot} \approx 0.2$
  to enable a search for \ovi\ absorption.   

\begin{figure*}[ht]
\epsscale{0.8}
\plotone{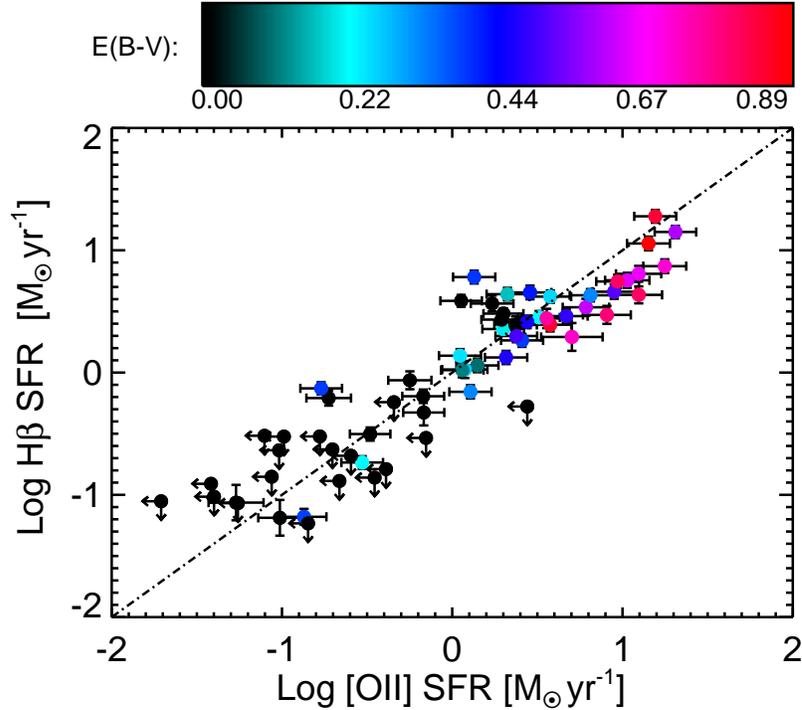}
\caption{  [OII] vs H$\beta$ SFR, where the color represents the
  magnitude of the E(B-V) Balmer correction derived from the observed
  H$\alpha$ to H$\beta$ ratio. Black points indicate that no Balmer
  correction was made because we did not detect H$\alpha$ or
  H$\gamma$. Those galaxies with the highest SFRs also tend to have
  the highest Balmer corrections. Upper limits to the SFRs are shown
  with left-facing and downward-facing arrows. The galaxies at high
  SFR tend to deviate from the one-to-one line because of a probable
  overestimation of the reddening at [OII]. \label{fig:sfrcomp}} 
\end{figure*}
  
\subsection{Spectral Measurements}

A galaxy spectrum dominated by active star formation will exhibit
emission lines that indicate its level of metal-enrichment and current
star formation rate (SFR). Here we describe the initial line
measurements and corrections made to the emission-line galaxies in our
sample. For non-emission-line galaxies in which  the spectrum is
dominated by stellar continuum and absorption, we describe the measurements that
allow us to obtain upper limits on the SFR.  

We developed an IDL-based
graphical user interface program (bundled within
XIDL\footnote{http://www.ucolick.org/$\sim$xavier/IDL/index.html}) named
$gal\_fit\_emiss$, to fit spectral features with Gaussian profiles
and/or boxcar fits that give integrated line fluxes and associated
photon noise errors. We compute the line FWHMs using
$galx\_fit\_emiss$, where the line Gaussian FWHM (km/s) = c
$\times$ FWHM (\AA) / $\lambda_{line}$ (\AA). The average line FWHMs
are 280 (blue) and 200 (red) km/s.  These values are dominated by the spectral resolution as the lines are unresolved.  For Balmer emission-lines, we
minimize contamination from underlying stellar absorption
by fitting the continuum in the trough of detectable absorption. The
overall effect on the line flux of the Balmer absorption ranges from
10\% to 60\% for the H$\beta$ emission line (when absorption is
apparent).  

We apply a correction for interstellar reddening to all line
measurements from the observed H$\gamma$ to H$\beta$  and H$\alpha$ to
H$\beta$ ratios for case B recombination where H$\gamma$/H$\beta$ =
0.459 and H$\alpha$/H$\beta$ = 2.86 at an effective temperature of
10,000 K and electron density of 100 cm$^{-3}$ \citep{hummer87}. We
use a reddening function normalized at H$\beta$ from the Galactic
reddening law of \cite{ccm} assuming R$_{v}$ = A$_{v}$/E(B$-$V) =
3.1. We do not apply a correction for internal interstellar reddening when we
cannot measure the line fluxes of at least 2 Balmer emission lines. We
tabulate the line fluxes and the E(B-V)$_{Balmer}$ in Tables
\ref{tab:gallines} and \ref{tab:galprops}.   


The error in the final reddening-corrected line flux measurement is
due to three primary factors:  (1) the photon noise, which is the least significant source of error, (2) the error in the Balmer
correction, which is based on the uncertainty in the flux ratio used
to calculate the Balmer correction and the intrinsic error ($\approx
10\%$) in the assumed constants (0.459, 2.86),
and (3) the error in the slit correction flux scale factor, assumed to be
10\%. The dominant error term is the Balmer correction, especially for
galaxies that turn out to be very dusty. The average reddening is E(B-V) = 0.3 for
the galaxies in our sample where we measured the Balmer
decrement.

\subsection{Star Formation Rates}

  \label{sec:sfr}

Once we correct the galaxy spectrum for the Balmer decrement, we
calculate a current SFR using the Balmer emission lines H$\alpha$ and
H$\beta$ and the [OII] $\lambda$$\lambda$ 3727 doublet.  For the
former, we use the calibration of \cite{kennicutt98} where SFR
[M$_{\odot}$ yr$^{-1}$] $=$ 7.9 $\times$ 10$^{-42}$ L$_{H\alpha}$
[ergs s$^{-1}$]. Balmer emission lines are the most straightforward of
emission-line SFR indicators because their luminosity directly traces
the ionizing stellar populations. The \cite{kennicutt98} H$\alpha$ SFR
calibration is derived from stellar population synthesis models that
assume a Salpeter IMF \citep{salpeter55} and solar metallicity.  When
H$\alpha$ is not observed in a spectrum, we use the same
\cite{kennicutt98} SFR calibration divided by a factor of 2.86 for the
H$\beta$ emission line. At an effective temperature of 10,000 K and
electron density of 100 cm$^{-3}$ for Case B recombination, this
factor of 2.86 is the intrinsic ratio of  H$\alpha$/H$\beta$
\citep{hummer87}. Thus, the SFRs derived from H$\alpha$ and H$\beta$
are identical when we calculate a dust correction that gives
H$\alpha$/H$\beta$ = 2.86.  

 Additionally, we tabulate [OII] SFRs using equation 4 of
 \cite{kewley04}, where SFR [M$_{\odot}$ yr$^{-1}$] $=$ 6.58 $\times$
 10$^{-42}$ L$_{[OII]}$ [ergs s$^{-1}$]. [OII] SFR indicators are more
 complicated than Balmer emission line indicators because [OII] is
 affected by reddening, ionization properties, stellar absorption, and
 metallicity. Ideally, we would adopt the [OII] SFR indicator that
 contains a correction for oxygen abundance, but not all of our
 galaxies have emission lines or wavelength coverage that permit a
 metallicity estimate. Figure \ref{fig:sfrcomp} compares [OII] and
 Balmer SFRs, and shows that they correlate well, but that there is a
 large scatter of $\sim$ 0.3 dex. This plot also shows the internal
 reddening correction in $E(B-V)$ for every galaxy in which we can
 measure it. The galaxies that deviate from the one-to-one line at
 high SFR are likely to suffer from an overestimation of the reddening
 at [OII]. The corrections at H$\alpha$ are small ($\sim$3\%) such that uncertainties in $E(B-V)$ and the extinction curve do not have a significant impact on the Balmer-derived SFRs. 
 
 When a galaxy's spectrum contains no emission lines, we measure an
 upper limit to the SFR by measuring the boxcar noise at the positions
 of [OII], H$\beta$ and H$\alpha$. We use 3$\sigma$ line flux limits
 as our SFR upper limits in these cases, approximately 1/3  of our
 sample.  We adopt conservative 3$\sigma$ limits to the SFRs since in these galaxies we are unable to make a correction for dust. There is a set of red galaxies where [OII] emission is present
 yet we have a strict limit on Balmer emission.  This is a somewhat
 common occurance in galaxy surveys and has been attributed to AGN
 activity \citep[e.g.][]{konidaris07}.  In these cases, we measure the
 [OII] line-flue but attribute an upper limit to the inferred SFR.
 These are generally much higher
 than the Balmer SFR upper limit for the same galaxies. In
 Table \ref{tab:galprops} we list SFRs, their errors, and mark the
 upper limits.  
\begin{figure*}[hpt]
\epsscale{1.0}
\plotone{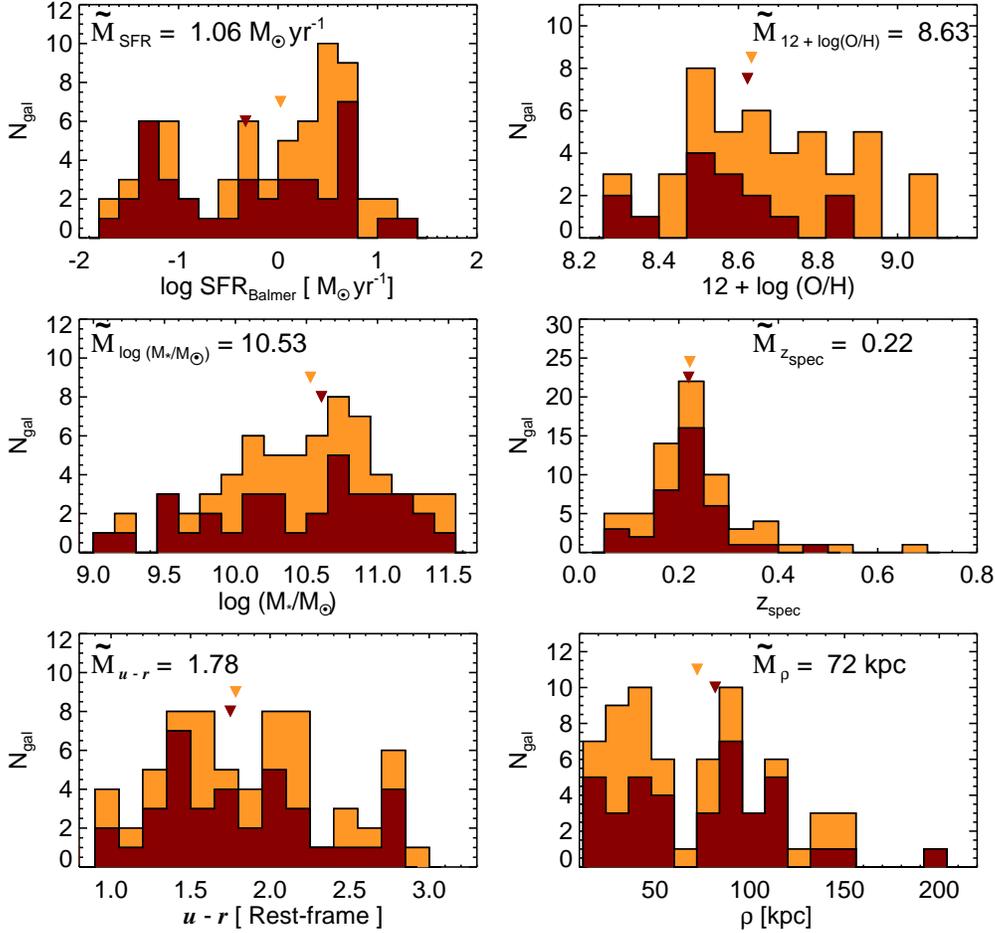}
\caption{   Distribution of several key derived galaxy properties for the entire sample of galaxies (light shade) and for target galaxies (dark shade). The median of each full distribution is marked with a light yellow, filled triangle, and labeled. The medians of the target  galaxy distributions are marked with a corresponding dark red filled triangle. \label{fig:hist}}
\end{figure*}
\subsection{Metallicity Determination}
\label{sec:abun}
When the necessary emission lines are present, we use two separate strong-line methods of determining the oxygen abundances for our foreground galaxies : the R$_{23}$ = ([OII] $\lambda$3727 + [OIII] $\lambda\lambda$4959, 5007) / H$\beta$ \citep{pagel79} calibration of McGaugh (1991; henceforth M91\nocite{mcgaugh91}), and the N2 index, the [NII]$\lambda$6583/H$\alpha$ ratio, based on the calibration of Pettini and Pagel (2004;  henceforth PP04\nocite{pettini04}). Below, we describe each method in detail and discuss the systematic uncertainties. 

The M91 R23 oxygen abundance has a relative error of  $\sim$0.15 dex over a wide range of abundances, but exhibits a well-known degeneracy, with a turn-over in the relation at Z $\sim$ 0.3 Z$_{\odot}$ (12$+$log(O/H) $\sim$ 8.35). The oxygen abundance is given by the following two analytic expressions for lower and upper branches \citep{kobulnicky99}: 
\begin{equation}
\begin{aligned}
12 + \log ({\rm O/H})_{lower} = &12   - 4.944 + 0.767x + 0.602{x}^2\\
&- y(0.29 + 0.332x - 0.331{x}^2), 
\end{aligned}
\end{equation}
\begin{equation}
\begin{aligned}
12 + \log ({\rm O/H})_{upper} &= 12   - 2.939 - 0.2x - 0.237x^{2}\\
& - 0.305x^{3} - 0.283x^{4} - y(0.0047\\
&- 0.0221x - 0.102x^{2} - 0.0817x^{3}\\
& - 0.00717x^{4}),
\end{aligned} 
\end{equation}
where x $=$ log R$_{23}$ and y $=$ log ([OIII] $\lambda$ 4959 + [OIII] $\lambda$ 5007)/ [OII] $\lambda$ $\lambda$ 3727. 

The most robust way to place a galaxy on the upper or lower branch of
the R23 relation is to use the [NII] $\lambda$6583 to [OII]
$\lambda$$\lambda$3727 ratio. When log [NII]/[OII] $<$ $-$1.0,  it
lies on the lower metallicity branch of the R23 relation (M91), and
correspondingly, when log [NII]/[OII] $>$ $-$1.0, an upper-branch
metallicity results \citep{kewley08}. The line flux ratio [NII]/H$\alpha$ (the N2 index) provides an
alternative method. If log N2 $<$ -1.3 ([NII]/H$\alpha$ $<$ 0.05)
there is a high degree of certainty that the oxygen abundance is on
the lower branch of the R23 relation. Whereas if log N2 $> -1.1$
([NII]/H$\alpha$ $>$ 0.08) the oxygen abundance is on the upper
branch. Between $-1.1$ and $-1.3$, the N2 index does not accurately
discriminate between upper and lower branches of R23 because the
oxygen abundance is likely to be very close to the turnover at 12 $+$
log (O/H) $=$ 8.3.  

The benefit to using the N2  index over log [NII]/[OII]  is that the former involves two lines in close wavelength proximity ($\lambda$ [NII] = 6583 \AA; H$\alpha$ = 6563 \AA)  such that the flux calibration and reddening correction have little to no impact on the resultant line ratio. The N2 index itself is sensitive to the metallicity to within 0.35 dex accuracy at a 95\% confidence level up to 12 + Log(O/H) = 8.8 (Pettini and Pagel 2004, henceforth PP04\nocite{pettini04}). We use the following expression from PP04 to calculate the oxygen abundance from the N2 index: 

\begin{equation}
12 + log({\rm O/H}) = 9.37 + 2.03 \times N2 + 1.26 \times N2^{2} + 0.32 \times N2^{3}, 
\end{equation}
where N2 $=$ log ([NII] $\lambda$6584/H$\alpha$). The PP04 calibration
is valid for $-2.5 < N2 < -0.3$, or 7.20 $<$ 12$+$ log (O/H) $<$
8.95. The benefits of this strong-line abundance indicator are that it
is monotonic with (O/H) and it does not require precise flux calibration or
a reddening correction due to the proximity in wavelength of [NII]
$\lambda$6583 and H$\alpha$.  The primary drawback is the large
uncertainty in the calibration (0.35 dex), and the limited range over
which it is useful. We tabulate PP04 oxygen abundances in Table
\ref{tab:galprops}, and note that there are several galaxies  with N2
indices $> -0.3$ which give values of 12$+$ log (O/H) outside the
valid range. We show these values of the PP04 abundance as lower
limits where  12 $+$ log (O/H) $>$ 8.95.  

\begin{figure*}[hpt]
\epsscale{1.0}
\plotone{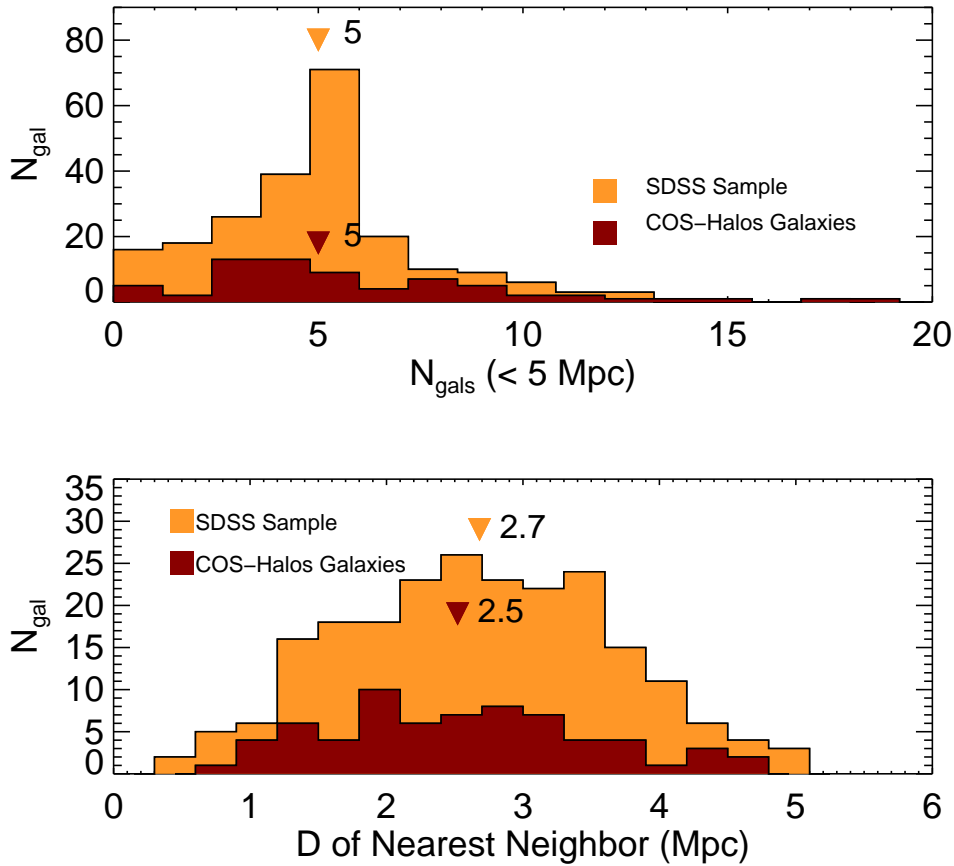}
\caption{Two histograms that show the number of galaxies within 5 Mpc of a given COS-halo galaxy (top) and the distance to the nearest neighboring galaxy in Mpc (bottom). In both plots the median values are marked for a randomly selected sample of SDSS galaxies in the same redshift range (light shade) and the COS-Halos galaxies (dark shade).  \label{fig:envirohist}}
\end{figure*}

\begin{figure*}[ht]
\epsscale{0.80}
\plotone{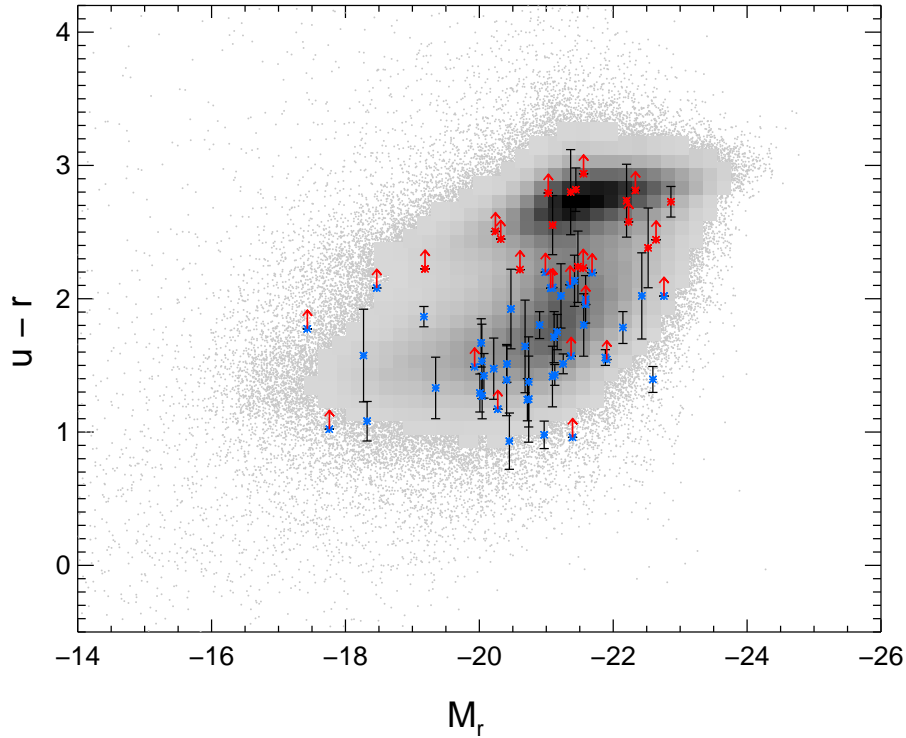}
\caption{  A color-magnitude plot for the COS-Halos galaxies, where blue galaxies ($u-r$ $<$  2.2) are plotted as blue asterisks and red galaxies ($u-r$ $\geq$  2.2)  are shown as red asterisks.  Error bars for the galaxy k-corrected colors are shown unless a red, upward-facing arrow indicates a u$-$r lower limit as discussed in Section \ref{sec:mass}. The shaded histogram is shown for local SDSS galaxies, and grey points are plotted when the number of SDSS galaxies is less than 100.   \label{fig:cmd}}
\end{figure*}

  \begin{figure*}[hpt]
\epsscale{0.85}
\plotone{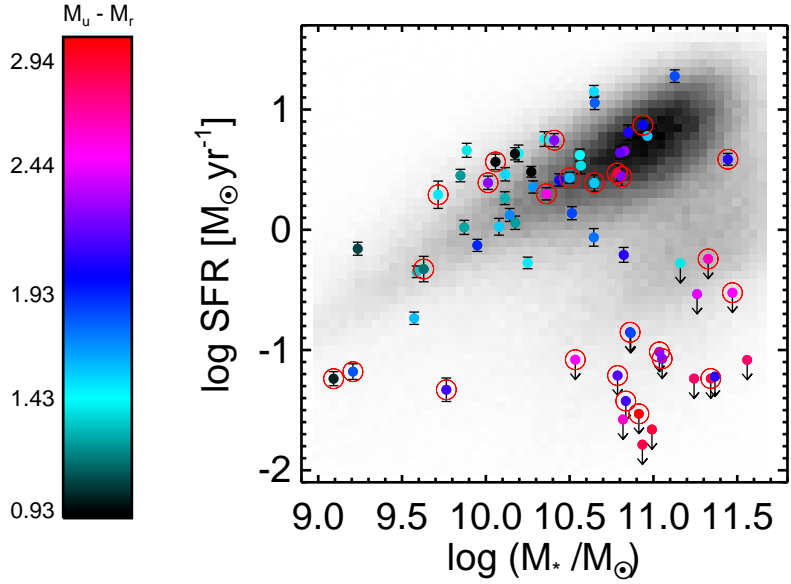}
\caption{Log SFR measured from either H$\alpha$ or H$\beta$ emission versus Log M$_{*}$.  Points in this figure are color-coded for galaxy u$-$r color, where red open circles indicate a u$-$r lower limit as discussed in \ref{sec:mass}. The number density distribution of $\sim$10$^{5}$ SDSS star-forming galaxies is shown in grayscale, for reference. The SFRs and stellar masses come from the MPA Value Added Catalogs, and are based on the comprehensive studies by \cite{brinchmann04}, for the SFR (median values are corrected to a Salpeter IMF, to match the calibration we use in this study), and by  \citep{kauffmann03}, for the stellar masses.  \label{fig:sfrmass}}
\end{figure*}

\begin{figure*}[hpt]
\epsscale{0.6}
\plotone{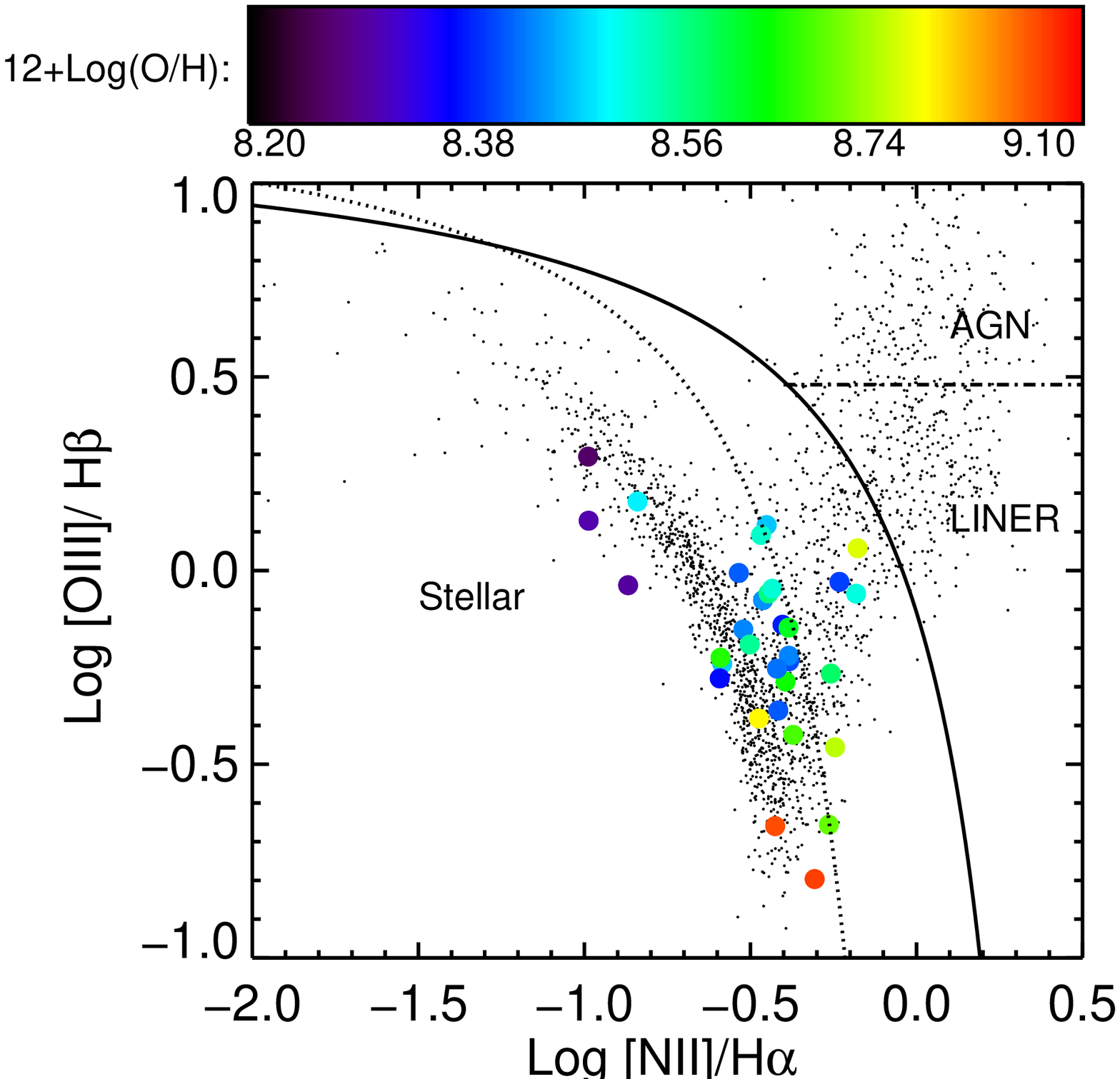}
\caption{ A BPT diagram \citep{baldwin81} showing the nominal flux ratios (log [OIII] $\lambda$5007/ H$\beta$ vs. log [NII] $\lambda$6584 / H$\alpha$) for a subset (32 of 67) of the entire sample of galaxies having available line measurements. The colors of the points correspond to galaxy oxygen abundances derived from the McGaugh 1991 calibration of the R23 relation. The black dots correspond to a random selection of galaxies from SDSS DR7, and roughly demarcate the full range of flux ratios observed for local galaxies. The two curves are two different calibrations (solid: Kewley et al. 2001; dashed: Kauffmann et al. 2003) that attempt to separate emission from ionized nebulae (star formation) from emission due to other processes (AGN, LINERs). The horizontal line separates AGN from LINERs.  \citep{veilleux87} \label{fig:diagnostic}}
\end{figure*}

In the cases for which we are unable to measure a Balmer decrement
(e.g. H$\beta$ falls in the dichroic), and the few cases for which we
do not trust the absolute flux calibration for red-blue side matching
(large uncertainty in SDSS apparent magnitudes), we prefer the N2
index to log [NII]/[OII] for breaking the degeneracy of the R23
relation. Furthermore, when H$\alpha$ and [NII] lie outside the
observed wavelength range for the galaxy, we cannot properly break the
degeneracy of the R23 relation (7 emission-line galaxies). Fortunately, there is a well-known
global relation between galaxy mass and metallicity, the
mass-metallcity relation \citep{skillman89, tremonti04}, that enables
us to make an informed guess as to R23 branch. The majority (66 of 68)
of our galaxies have stellar masses $>$ 10$^{9}\mmsun$, making it more
likely that they are on the upper branch of the R23 relation than the
lower branch. In Section \ref{sec:results}, we further discuss the
mass-metallicity relation of our sample of foreground galaxies.   


\cite{mcgaugh91} reports several different values of the systematic
error associated with this strong-line calibration depending on the
resultant oxygen abundance: 0.1 dex for the upper branch, 0.05 dex for
the lower branch, and 0.20 dex within 0.1 dex of the R23
turnover. Additional, unaccounted for sources of error arise from HII
region age-effects (M91 is calibrated for zero-age HII regions) and
geometrical effects \citep{stasinska96, ercolano07}. The overall
impact of these effects increases the systematic error in this method
to 0.1 $-$ 0.3 dex, on average, regardless of branch, and in the
``worst-case" scenario. We adopt $\pm$0.15 dex as an average systematic
error in our abundance measurements, unless a value is within 0.1 dex
of the turnover, where it is then estimated to be $\pm$0.2 dex. Neither the
M91 nor PP04 calibration indicates an oxygen abundance on an absolute
scale better than to a factor of $\sim$0.3\,dex, and there are well-known
systematic offsets between the two methods \citep{kewley08}.  

\section{Results: Global Properties of the Galaxies} 
\label{sec:results}
 \nocite{kewley01, kauffmann03}
 As described in Section 1, the data and analysis presented in
the proceeding sections are associated with a galaxy sample selected
from the SDSS for a targeted study of 
gas in the halos of $L \approx L^*$ galaxies at $z \sim 0.2$.
This blind survey was designed to sample galaxies with a
range of stellar mass, star-formation rate, and color.
In this final section, we describe the distribution of galaxy
characteristics for the sample,  
provide global context for their properties, 
and highlight differences (if any) from the general
low-$z$ population.  

 The properties of the galaxies in our sample are summarized by the
  histograms in Figure \ref{fig:hist}, where the full galaxy sample
  (targets + bonus) is shown by the light-yellow shade histograms and the
  target galaxies are shown by the dark-orange shade histograms. 
  The distribution of SFRs shows a bimodality, separated at
  $\sim$0.1 M$_{\odot}$ yr$^{-1}$. The majority of SFRs below this
  value are upper limits, i.e. the lower of the two three-sigma Balmer
  SFR limits. This apparent bimodality, then, is more a reflection of the
  sensitivity limit of our spectral data than it is a sign of any
  physical bimodality of SFRs in our sample.  

  As we discussed in
  Section \ref{sec:mass}, our original sample criteria select against lower-mass galaxies
  compared to the universal distribution.
  These same selection effects also result in a metallicity
  distribution that is lacking in the
  lowest-metallicity galaxies (i.e. dwarfs), as expected from the
  mass-metallciity relation.  As expected and desired, our selection of galaxies
  to have $zphot$ between 0.11 and 0.4 leads to a distribution in the
  spectroscopic redshifts that is clustered around a median value of
  $\sim$ 0.2. As seen in Figure \ref{fig:zphot}, the SDSS-based
  $zphot$ is occasionally highly skewed, and there are several
  galaxies in our sample that we found to have redshifts $<$ 0.11 or
  $>$ z$_{\rm QSO}$, making their OVI lines unobservable with the COS
  spectrograph. The median impact parameter
  for our sample is 118 kpc in the galaxy rest frame. Finally,
  k-corrected $u - r$ colors show a distribution between 1 and 3, with
  a median of 1.8 magnitudes. 

To examine the environments of the COS-Halos galaxies, we first searched the maxBCG galaxy cluster catalog \citep{maxbcg} for any likely matches. The following galaxies turn out to lie in or near a maxBCG galaxy cluster: J0928+6025: 110\_35 (7.95 Mpc from cluster center), 129\_19 (7.38 Mpc from cluster center), and 187\_15 (9.08 Mpc from cluster center)); J1157-0022: 359\_16 (12.35 Mpc from cluster center); and J1514+3620: 287\_14 (19.3 Mpc from the cluster center). Furthermore, we compare the  neighborhood of the COS-Halos galaxies to that of a random selection of $\sim$500 SDSS galaxies of similar luminosities and in the same redshift range. Figure \ref{fig:envirohist} shows the results of this comparison. The distances between galaxies are comoving distances computed such that D$_{tot}$ = $\sqrt(D_{CM, z1}^{2} + D_{CM, z2}^{2} - 2D_{CM, z1}D_{CM, z2}cos\theta)$. In this equation, D$_{tot}$ is the comoving distance from one galaxy at redshift z1 to another galaxy at redshift z2 with a projected angular separation, $\theta$, in radians. Figure \ref{fig:envirohist} shows the number of galaxies within 5 Mpc of a given galaxy and the distance to the nearest neighboring galaxy in Mpc. When environment is assessed in this manner, we find essentially no differences between our sample and the control SDSS sample. 

We show a color-magnitude diagram for the COS-Halos galaxies and numerous SDSS galaxies in Figure \ref{fig:cmd}. The k-corrected colors and absolute magnitudes of the SDSS galaxies come from the NYU Value-Added Galaxy Catalog \citep{blanton05}, and are further corrected to account for our adopted cosmology (i.e. a factor of 5$log$h). Although the COS-Halos galaxies fall within the main locus of SDSS galaxies, blue galaxies are somewhat over-represented in our target galaxy sample. One goal of the COS-Halos project is to examine the relation of galaxy star-forming properties to halo properties, and we originally selected a blue-biased galaxy sample to probe the full range of galaxy star-forming properties.

 A galaxy's SFR is anti-correlated with its stellar mass, a trend we reproduce in Figure \ref{fig:sfrmass}. Points in this figure are color-coded for galaxy u$-$r color  and show that the redder galaxies with SFR upper limits in this sample are on average more massive than bluer SF galaxies. The red open circles indicate a u$-$r lower limit as discussed in \ref{sec:mass}. The number density distribution of $\sim$10$^{5}$ SDSS star-forming galaxies is shown in grayscale, for reference. The SFRs and stellar masses come from the MPA Value Added Catalogs \footnote{http://www.mpa-garching.mpg.de/SDSS/DR4/Data/sfr\_catalogue.html}, and are based on the comprehensive studies by \cite{brinchmann04}, for the SFR (median values are corrected to a Salpeter IMF, to match the calibration we use in this study), and by  \cite{kauffmann03}, for the stellar masses. In this parameter space, the COS-Halos galaxies appear to trace the bimodal distribution of SDSS galaxies such that red, massive galaxies with very low SFRs separate cleanly from bluer galaxies with higher SFRs.  

We plot the log [OIII] $\lambda$5007/ H$\beta$ vs. log [NII]
$\lambda$6584 / H$\alpha$ line flux ratios in Figure
\ref{fig:diagnostic}, in a BPT diagram \citep{baldwin81} useful for
separating emission due to ionized nebulae (star formation) from
emission due to other processes (AGN, LINERs). The star-forming
sequence is delineated by two curves: the dotted line
\citep{kauffmann03} separates purely star-forming galaxies (left) from
other types of active galaxies (right), and the solid line
\citep{kewley01} demarcates galaxies that have emission from combined
sources, including star formation, from pure AGN. Every emission-line
galaxy in our sample is associated with at least some star formation,
with 5/30 exhibiting combined emission. For these 5 galaxies,
metallicities and star formation rates have more complicated
interpretations since the strength of emission lines will not directly
correlate with either quantity in an AGN spectrum. These galaxies are
classified as ``combined SF/AGN" in the galaxy type column of Table
\ref{tab:galprops}.

\begin{figure}[h!]
\epsscale{1.35}
\begin{centering}
\plotone{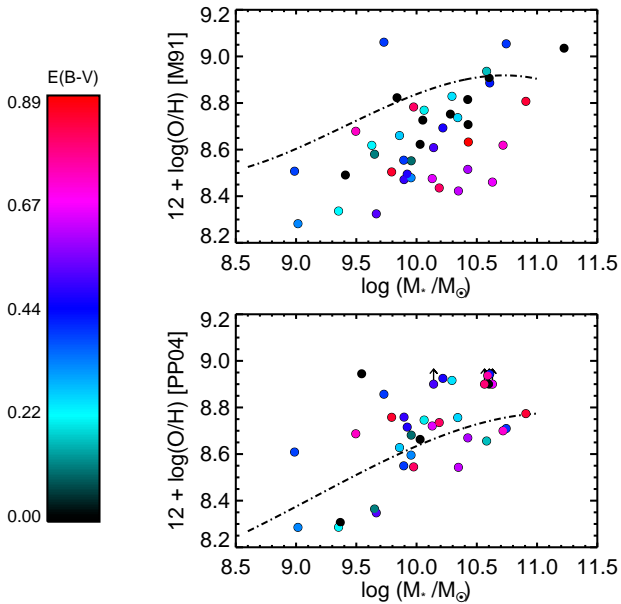}
\end{centering}
\caption{ The mass-metallicity relation for a subset of galaxies in
  our sample having metallicity estimates based on the M91 calibration
  of the R23 relation (top panel) and based on the PP04 calibration of
  the N2 index (lower panel). Stellar masses are determined from
  Michael Blanton's $k-correct$ \citep{blanton07}, and corrected for a
  factor of $5logh$ where $h = 0.72$ in our adopted cosmology. The
  dashed-dotted curves come from fits to SDSS data from
  \cite{kewley08}, for both the M91 (top) and PP04 (bottom)
  calibrations. \label{fig:massmetal}} 
\end{figure}

In Figure \ref{fig:massmetal} we show the mass-metallicity relation
for the emission line galaxies in our sample using both M91 and PP04
oxygen abundance calibrations. Generally, we reproduce the well-known
trend that more massive galaxies tend to contain more
metals. Comparing our relations with the calibration-dependent fits
to SDSS data of \cite{kewley08}, we see that our M91 mass-metallicity
relation is discrepant from the fit to SDSS data of \cite{kewley08}, while the PP04 fit is
better. It is unclear what is causing this discrepancy in the M91
calibration mass-metallicity relation.  The offset is opposite what we would expect if reddening corrections are systematically underestimated (due to not being able to make any reddening corrections for several galaxies because of lacking necessary emission lines). However, even if reddening is overestimated for many of the galaxies, its effect would be, at most, on the level of 0.1 dex. Furthermore, we do not find a significant offset between M91 and PP04 oxygen abundances for our sample of galaxies, as do \cite{kewley08}. The typical scatter in the mass-metallicity relation (0.3 dex) is reflected in these plots.

\section{Summary}
\label{sec:rogues:sum}
In this paper, we describe the details of the optical observations and spectral analyses done as part of the COS-Halos survey (Tumlinson et al. 2011).    The high signal-to-noise optical spectra for 67 galaxies from Keck LRIS and Magellan MagE presented here are essential to the COS-Halos survey that aims to explore the variations of halo gas properties with galaxy properties. We determine and  tabulate galaxy spectroscopic redshifts accurate to $\pm$ 30 km s$^{-1}$,  impact parameters, rest-frame colors, stellar masses, total star formation rates, and gas-phase interstellar medium oxygen abundances.

The COS-Halos target galaxy sample was pre-selected to span redshifts of 0.1 $-$ 0.3, stellar masses log(M$_{*}$/M$_{\odot}$) = 9.5 $-$ 11.5, and be located $<$ 160 kpc from a background QSO sightline. These criteria bias the galaxies to have higher than average galaxy masses (and, by extension, higher than average metallicities), and may further select against galaxies with spectra dominated by AGN (see Figure \ref{fig:diagnostic}). Within the pre-selection criteria, the COS-Halos galaxies are well-sampled with respect to mass and star formation, with 2/3 of the galaxies being dominated by ongoing star-formation. A wide range of SFRs (0.01 $-$ 20 M$_{\odot}$ yr $^{-1}$) will allow us to investigate the connection between galaxy bimodality and halo gas. Although three of the COS-Halos target galaxies were found to be part of Galaxy clusters, we do not find that the environments of the galaxies are on average significantly different from those of the general low-redshift, L* galaxy population. Through this analysis, we are able to reproduce well-known correlations between galaxy metallicity and mass, galaxy global SFRs and galaxy mass, and find no significant deviations.  In total, the COS-Halos galaxy sample is representative of a normal set of L$\sim$L* galaxies at z$\sim$ 0.2.  Subsequent analyses using the COS-Halos survey data will rely on the galaxy properties determined in this work.

 \section{Acknowledgements}
   
  Support for program GO11598 was provided by NASA through a grant from the Space Telescope Science Institute, which is operated by the Association of Universities for Research in Astronomy, Inc., under NASA contract NAS 5-26555.  Much of the data presented herein were obtained at the W.M. Keck Observatory, which is operated as a scientific partnership among the California Institute of Technology, the University of California and the National Aeronautics and Space Administration. The Observatory was made possible by the generous financial support of the W.M. Keck Foundation. The authors wish to recognize and acknowledge the very significant cultural role and
reverence that the summit of Mauna Kea has always
had within the indigenous Hawaiian community. We are
most fortunate to have the opportunity to conduct observations from this mountain.

    {{\it Facilities:} \facility{Keck: LRIS} \facility{Magellan: MagE}}

\bibliography{opticalspec}
\bibliographystyle{apj}

\clearpage
\begin{table*}[hpt] \centering \scriptsize
\begin{tabular*}{0.9\textwidth}{@{\extracolsep{\fill}}llcccccllcc}
\hline
Field & ID & RA & Dec & t$_{blue}$ & t$_{red}$ & m$_{g}$ & m$_{i}$ & F$_{B}$ & F$_{R}$ \\
\hline
&&&&Targets: \\
\hline
\hline
J0042-1037&358\_9&00:42:22.27&-10:37:35.2&2008-10-05&2 $\times$  900&2$ \times$  900&20.57$\pm$0.055&19.57$\pm$0.043& 2.91& 2.50\\
J0226+0015&268\_22&02:26:12.98&+00:15:29.1&2008-10-04&2 $\times$  900&2$ \times$  875&20.79$\pm$0.048&18.93$\pm$0.020& 1.31& 1.27\\
J0401-0540&67\_24&04:01:50.48&-05:40:47.0&2008-10-05&2 $\times$  900&2$ \times$  900&20.11$\pm$0.048&19.19$\pm$0.050& 2.17& 3.50\\
J0803+4332&306\_20&08:03:57.74&+43:33:09.9&2011-05-01&  600&2$ \times$  270&19.92$\pm$0.030&17.95$\pm$0.012& 5.90& 2.76\\
J0820+2334&260\_17&08:20:22.99&+23:34:47.4&2009-03-24&  540&  540&20.61$\pm$0.061&19.40$\pm$0.050& 6.89& 9.61\\
J0910+1014&35\_14&09:10:30.30&+10:14:25.0&2009-03-24&  600&2$ \times$  600&21.11$\pm$0.075&19.27$\pm$0.029& 3.08& 1.89\\
J0914+2823&41\_27&09:14:41.75&+28:23:51.3&2009-03-24&  600&  600&20.29$\pm$0.032&19.50$\pm$0.033& 1.71& 1.77\\
J0925+4004&193\_25&09:25:54.23&+40:03:50.1&2009-03-24&  300&  300&20.25$\pm$0.026&19.04$\pm$0.018& 1.41& 1.18\\
J0928+6025&110\_35&09:28:42.46&+60:25:08.7&2010-03-25&  900&2$ \times$  410&19.37$\pm$0.018&17.92$\pm$0.011& 1.52& 1.48\\
J0943+0531&106\_34&09:43:33.78&+05:31:22.2&2010-03-25&2 $\times$  600&  410&20.00$\pm$0.027&18.53$\pm$0.015& 3.21& 1.30\\
J0950+4831&177\_27&09:50:00.86&+48:31:02.2&2010-03-25&  900&2$ \times$  430&19.28$\pm$0.017&17.60$\pm$0.009& 1.86& 1.78\\
J1009+0713&204\_17&10:09:01.58&+07:13:28.0&2010-03-25&  900&2$ \times$  410&20.49$\pm$0.039&19.61$\pm$0.038& 2.33& 1.98\\
J1016+4706&274\_6&10:16:22.02&+47:06:43.7&2010-04-05&  800&2$ \times$  360&21.10$\pm$0.065&20.09$\pm$0.052& 1.45& 1.16\\
J1022+0132&337\_29&10:22:18.22&+01:32:45.4&2010-03-25&  630&  630&20.38$\pm$0.112&19.95$\pm$0.176& 6.53& 3.99\\
J1112+3539&236\_14&11:12:38.16&+35:39:20.4&2010-03-25&  900&2$ \times$  410&20.15$\pm$0.031&19.13$\pm$0.025& 4.05& 1.81\\
J1133+0327&110\_5&11:33:28.08&+03:27:17.5&2010-03-25&  900&2$ \times$  410&19.12$\pm$0.023&17.59$\pm$0.014& 2.86& 2.50\\
J1157-0022&230\_7&11:57:58.36&-00:22:25.4&2010-03-25&  900&2$ \times$  410&...&...&...&...\\
J1220+3853&225\_38&12:20:32.82&+38:52:49.7&2010-03-25&  900&2$ \times$  410&20.73$\pm$0.056&19.14$\pm$0.024& 2.06& 1.34\\
J1233+4758&50\_39&12:33:38.01&+47:58:25.5&2010-04-05&  800&  360&19.57$\pm$0.019&18.60$\pm$0.019& 8.49& 2.45\\
J1233-0031&242\_15&12:33:03.17&-00:31:41.2&2010-04-05&  800&2$ \times$  360&21.00$\pm$0.076&20.00$\pm$0.066& 9.28& 6.48\\
J1241+5721&199\_6&12:41:53.76&+57:21:01.4&2010-03-25&  900&  410&20.76$\pm$0.036&19.72$\pm$0.031& 1.33& 1.39\\
J1245+3356&236\_36&12:45:08.88&+33:55:50.1&2010-03-25&  620&  410&20.39$\pm$0.039&19.54$\pm$0.032& 1.20& 1.29\\
J1322+4645&349\_11&13:22:22.46&+46:45:46.1&2010-03-25&  900&2$ \times$  410&20.16$\pm$0.026&18.60$\pm$0.017& 1.21& 1.26\\
J1330+2813&289\_28&13:30:43.13&+28:13:30.4&2010-04-05&  800&2$ \times$  360&20.77$\pm$0.034&19.31$\pm$0.019& 1.36& 1.10\\
J1419+4207&132\_30&14:19:12.21&+42:07:26.5&2010-03-25&  900&2$ \times$  410&19.41$\pm$0.016&18.20$\pm$0.015& 1.56& 1.79\\
J1435+3604&68\_12&14:35:12.41&+36:04:41.5&2010-04-05&  800&2$ \times$  360&18.84$\pm$0.015&17.54$\pm$0.013& 9.38& 3.52\\
J1437+5045&317\_38&14:37:23.43&+50:46:23.5&2010-04-05&  800&2$ \times$  360&19.98$\pm$0.022&19.27$\pm$0.028& 1.65& 1.45\\
J1445+3428&232\_33&14:45:09.21&+34:28:05.3&2010-04-05&  800&2$ \times$  360&20.78$\pm$0.032&19.40$\pm$0.023& 1.51& 1.19\\
J1514+3619&287\_14&15:14:27.56&+36:20:02.0&2010-03-25&  900&2$ \times$  410&20.93$\pm$0.055&19.91$\pm$0.050& 1.34& 1.70\\
J1550+4001&197\_23&15:50:47.70&+40:01:22.6&2010-04-05&  800&2$ \times$  360&20.42$\pm$0.036&18.42$\pm$0.014& 1.84& 1.46\\
J1555+3628&88\_11&15:55:05.26&+36:28:48.4&2010-03-25&  900&2$ \times$  410&19.36$\pm$0.018&18.40$\pm$0.014& 1.64& 1.54\\
J1616+4154&327\_30&16:16:47.99&+41:54:41.3&2010-03-25&  820&  410&20.39$\pm$0.027&19.76$\pm$0.034& 1.69& 1.02\\
J1619+3342&113\_40&16:19:19.51&+33:42:22.8&2010-03-25&  600&  410&19.77$\pm$0.021&18.78$\pm$0.019& 1.11& 1.40\\
J2257+1340&270\_40&22:57:35.43&+13:40:45.3&2008-10-04&2 $\times$  600&2$ \times$  600&19.55$\pm$0.020&17.96$\pm$0.011& 1.0& 1.0\\
J2345-0059&356\_12&23:45:00.37&-00:59:23.9&2008-10-04&2 $\times$  900&2$ \times$  900&20.10$\pm$0.045&18.72$\pm$0.027& 1.30& 1.39\\
\hline
&&&&Bonus: \\
\hline
J0820+2334&242\_9&08:20:23.62&+23:34:46.1&2010-04-05&  800&2 $\times$  360&21.36$\pm$0.070&20.41$\pm$0.069& 1.15& 1.61\\
J0910+1014&34\_46&09:10:31.50&+10:14:51.1&2009-03-24&  600&2 $\times$  600&18.52$\pm$0.013&17.60$\pm$0.010& 2.15& 2.47\\
J0914+2823&41\_123&09:14:46.57&+28:25:03.9&2009-03-24&  600&  600&19.78$\pm$0.024&18.40$\pm$0.019& 3.93& 2.61\\
J0925+4004&196\_22&09:25:54.18&+40:03:53.4&2009-03-24&  300&  300&20.23$\pm$0.034&17.99$\pm$0.011& 1.56& 2.28\\
J0928+6025&129\_19&09:28:39.99&+60:25:08.9&2010-03-25&  900&2 $\times$  410&19.47$\pm$0.022&18.65$\pm$0.025& 1.91& 2.01\\
J0928+6025&187\_15&09:28:37.75&+60:25:06.3&2010-03-25&  900&2 $\times$  410&20.53$\pm$0.041&19.75$\pm$0.048& 2.70& 7.91\\
J0928+6025&188\_7&09:28:37.85&+60:25:14.3&2010-03-25&  900&2 $\times$  410&20.95$\pm$0.053&19.03$\pm$0.023& 2.85& 1.35\\
J0928+6025&90\_15&09:28:40.01&+60:25:21.0&2010-04-05&  800&2 $\times$  360&21.11$\pm$0.055&20.01$\pm$0.048& 1.39& 1.26\\
J0943+0531&216\_61&09:43:29.20&+05:30:41.8&2010-03-25&  900&2 $\times$  410&18.90$\pm$0.013&17.38$\pm$0.007& 1.34& 1.43\\
J0943+0531&227\_19&09:43:30.67&+05:31:18.1&2010-03-25&2 $\times$  900&2 $\times$  410&22.18$\pm$0.145&21.12$\pm$0.113& 1.95& 1.85\\
J0943+0531&29\_23&09:43:32.37&+05:31:52.0&2010-03-25&2 $\times$  900&2 $\times$  410&22.03$\pm$0.122&20.93$\pm$0.091& 2.18& 2.63\\
J1009+0713&170\_9&10:09:02.17&+07:13:34.6&2010-04-05&  800&2 $\times$  360&21.56$\pm$0.059&20.69$\pm$0.061& 1.46& 1.56\\
J1009+0713&86\_4&10:09:02.27&+07:13:43.9&2010-04-05&  800&2 $\times$  360&...&...&...&...\\
J1016+4706&359\_16&10:16:22.58&+47:06:59.4&2010-04-05&  800&2 $\times$  360&19.40$\pm$0.019&18.31$\pm$0.015& 3.03& 2.53\\
J1133+0327&164\_21&11:33:28.15&+03:26:59.1&2010-04-05&  800&2 $\times$  360&19.87$\pm$0.027&19.02$\pm$0.030& 2.02& 1.73\\
J1133+0327&203\_10&11:33:27.51&+03:27:09.6&2010-03-25&  900&2 $\times$  410&20.08$\pm$0.025&18.51$\pm$0.015& 1.0& 1.0\\
J1233+4758&94\_38&12:33:38.87&+47:57:57.6&2010-04-05&  800&  360&19.92$\pm$0.023&18.48$\pm$0.016& 1.99& 2.16\\
J1233-0031&168\_7&12:33:04.14&-00:31:40.5&2010-04-05&  800&2 $\times$  360&21.60$\pm$0.109&20.19$\pm$0.070& 3.18& 3.90\\
J1241+5721&208\_27&12:41:52.45&+57:20:43.7&2010-03-25&  900&  410&20.92$\pm$0.047&19.82$\pm$0.040& 1.92& 1.29\\
J1330+2813&83\_6&13:30:45.63&+28:13:22.3&2010-04-05&  800&2 $\times$  360&21.23$\pm$0.063&20.18$\pm$0.050& 1.61& 1.41\\
J1435+3604&126\_21&14:35:12.93&+36:04:25.0&2010-04-05&  800&2 $\times$  360&21.03$\pm$0.044&19.76$\pm$0.039& 2.05& 1.80\\
J1437+5045&24\_13&14:37:26.68&+50:46:07.4&2010-04-05&  800&2 $\times$  360&20.88$\pm$0.046&19.66$\pm$0.041& 1.65& 1.51\\
J1445+3428&231\_6&14:45:10.90&+34:28:21.7&2010-04-05&  800&2 $\times$  360&22.11$\pm$0.094&20.56$\pm$0.058& 4.25& 1.06\\
J1550+4001&97\_33&15:50:51.11&+40:01:41.0&2010-04-05&  800&2 $\times$  360&20.79$\pm$0.063&19.29$\pm$0.034& 4.53& 2.42\\
J2257+1340&230\_25&22:57:36.90&+13:40:29.3&2008-10-04&  600&  600&19.92$\pm$0.036&18.32$\pm$0.019& 2.38& 1.90\\
J2257+1340&238\_31&22:57:36.42&+13:40:29.3&2008-10-04&  600&  600&20.08$\pm$0.057&18.34$\pm$0.026& 2.52& 2.40\\
\hline
\hline
\end{tabular*}
\caption[Target and bonus galaxies observed with Keck LRIS] {  (1) SDSS Field Identifier
(2) Galaxy Identifier, where the first number is the position angle in degrees from the QSO
and the second number is the projected separation in arcseconds (impact parameter) from the QSO
(3
(4) Galaxy declination, in degrees, minutes, seconds
(5) The date of the observation in the form YYYY-MM-DD 
(6) \& (7) The exposure time in seconds, on the blue and red sides
(8) \& (9) SDSS G-band and I-band magnitudes, and associated errors. 
These quanties are used to perform the correction to an absolute flux scale 
(10) \& (11) The slit-correction flux scale factors for the red and blue sides 
 \label{tab:keck_obs}}
\end{table*}

\begin{table*}[htp] \centering \scriptsize
\begin{tabular*}{0.7\textwidth}{@{\extracolsep{\fill}}llcccclc}

\hline
Field &ID &RA&Dec &Date&t$_{exp}$&m$_{r}$&F$_{spec}$\\
\hline

&&&Targets: \\
\hline
\hline
J0935+0204&15\_28&09:35:18.66&+02:04:42.8&2011-03-28& 1200&19.37$\pm$0.022& 1.13\\
J1342-0053&157\_10&13:42:51.85&-00:53:54.2&2011-03-28&  900&18.48$\pm$0.010& 1.57\\
J1617+0638&253\_39&16:17:08.92&+06:38:22.2&2011-03-29&  600&16.56$\pm$0.005& 3.00\\
\hline
&&&Bonus: \\
\hline
J0910+1014&242\_34&09:10:27.70&+10:13:57.2&2011-03-29& 1000&18.26$\pm$0.014& 2.76\\
J1342-0053&304\_29&13:42:49.99&-00:53:29.0&2011-03-29& 1200&18.36$\pm$0.010& 1.70\\
J1342-0053&77\_10&13:42:52.23&-00:53:43.2&2011-03-28&  600&19.92$\pm$0.030& 2.87\\
\hline
\hline
\end{tabular*}
\caption{ Target and bonus galaxies observed with Magellan Mage:  (1) SDSS Field Identifier
(2) Galaxy Identifier, where the first number is the position angle in degrees from the QSO
and the second number is the projected separation in arcseconds (impact parameter) from the QSO
(3) Galaxy Right Ascension, in hours, minutes, seconds 
(4) Galaxy declination, in degrees, minutes, seconds
(5) The date of the observation in the form YYYY-MM-DD 
(6)  The exposure time in seconds
(7) SDSS r-band magnitude, and associated error. 
These quanties are used to perform the correction to an absolute flux scale 
(8)  The slit-correction flux scale factor based on a comparison to the SDSS r-band magnitude. \label{tab:mageobs}}  
\end{table*}

\begin{table*}[hp]\centering \scriptsize
\begin{tabular*}{0.9\textwidth}{@{\extracolsep{\fill}}llccccccc}

\hline
Field &ID & [OII] &H$\gamma$ & H$\beta$ &[OIII] & [OIII] &
H$\alpha$ & [NII]  \\
 && $\lambda$$\lambda$3727& &  &$\lambda$4959& $\lambda$5007 &
& $\lambda$6584\\
\hline
&&&&Targets: \\
\hline
\hline
J0042-1037&358\_9&   79.4$\pm$   1.0&    5.2$\pm$   0.4&   18.3$\pm$   0.7&    8.8$\pm$   0.6&   29.2$\pm$   0.8&   65.6$\pm$   0.8&   13.1$\pm$   0.5\\
J0226+0015&268\_22&$<$    2.4&...&$<$    2.7&...&    6.1$\pm$   0.6&$<$    3.4&...\\
J0401-0540&67\_24&  104.8$\pm$   1.1&    7.3$\pm$   0.7&   27.7$\pm$   1.7&   16.7$\pm$   1.4&   38.5$\pm$   1.4&   87.7$\pm$   2.6&   31.1$\pm$   2.6\\
J0803+4332&306\_20&$<$    8.8&...&$<$    5.1&...&...&$<$    4.1&...\\
J0820+2334&260\_17&   70.6$\pm$   8.0&...&$<$   13.7&...&   20.0$\pm$   4.6&   28.3$\pm$   5.7&   16.1$\pm$   5.5\\
J0910+1014&35\_14&$<$    6.6&...&$<$    3.1&...&...&...&...\\
J0914+2823&41\_27&  106.5$\pm$   2.4&   14.5$\pm$   1.8&   34.8$\pm$   2.2&   14.2$\pm$   2.2&   37.4$\pm$   2.4&...&...\\
J0925+4004&193\_25&   50.5$\pm$   2.9&   16.1$\pm$   2.3&   22.5$\pm$   2.6&...&   21.8$\pm$   2.5&...&...\\
J0928+6025&110\_35&$<$    9.9&...&...&...&...&$<$    5.6&...\\
J0935+0204&15\_28&$<$    3.1&...&$<$    2.2&...&...&$<$    6.9&...\\
J0943+0531&106\_34&   47.1$\pm$   5.1&...&   34.1$\pm$   3.2&...&   14.2$\pm$   3.4&  148.8$\pm$   2.7&   84.9$\pm$   2.5\\
J0950+4831&177\_27&$<$   14.6&...&$<$   11.1&...&...&$<$    6.1&   29.6$\pm$   2.3\\
J1009+0713&204\_17&   97.2$\pm$   3.9&    6.6$\pm$   3.1&   25.3$\pm$   2.9&...&   27.9$\pm$   2.7&  121.4$\pm$   2.2&   16.5$\pm$   1.8\\
J1016+4706&274\_6&   57.7$\pm$   1.0&    5.1$\pm$   0.8&   15.9$\pm$   0.7&    7.2$\pm$   0.6&   22.4$\pm$   0.8&   37.7$\pm$   0.7&   12.8$\pm$   0.5\\
J1022+0132&337\_29&   65.6$\pm$  16.8&...&$<$   30.7&...&...&   58.8$\pm$   5.4&...\\
J1112+3539&236\_14&   56.7$\pm$   5.3&...&   17.5$\pm$   2.5&...&    9.3$\pm$   2.4&   94.6$\pm$   1.8&   36.5$\pm$   1.7\\
J1133+0327&110\_5&$<$   15.9&...&$<$    8.4&...&...&...&...\\
J1157-0022&230\_7&$<$    4.4&...&$<$    5.8&...&...&...&...\\
J1220+3853&225\_38&$<$    6.6&...&$<$    2.7&...&...&$<$    3.6&...\\
J1233+4758&50\_39&$<$   21.6&...&$<$    5.0&...&...&...&...\\
J1233-0031&242\_15&   46.7$\pm$   4.7&...&   14.2$\pm$   1.8&...&   12.4$\pm$   2.7&...&...\\
J1241+5721&199\_6&   49.5$\pm$   2.3&    4.7$\pm$   1.5&   11.8$\pm$   2.3&...&    8.8$\pm$   2.1&   74.9$\pm$   2.2&   19.4$\pm$   1.6\\
J1245+3356&236\_36&  108.1$\pm$   2.8&    5.7$\pm$   2.0&   32.5$\pm$   2.6&...&   56.1$\pm$   2.3&  104.4$\pm$   2.3&   15.1$\pm$   2.4\\
J1322+4645&349\_11&   21.7$\pm$   2.2&...&   21.2$\pm$   1.6&...&   27.3$\pm$   1.4&   61.5$\pm$   1.8&   43.4$\pm$   1.6\\
J1330+2813&289\_28&   39.8$\pm$   1.0&    5.4$\pm$   0.6&   16.7$\pm$   0.7&...&   17.5$\pm$   0.9&   79.3$\pm$   1.0&   57.0$\pm$   0.8\\
J1342-0053&157\_10&   25.2$\pm$   3.8&...&   50.4$\pm$   3.1&...&   13.1$\pm$   1.7&  214.3$\pm$   4.9&   97.2$\pm$   4.1\\
J1419+4207&132\_30&   52.8$\pm$   2.1&    9.4$\pm$   1.9&   31.8$\pm$   2.3&    4.7$\pm$   2.1&   13.6$\pm$   2.5&...&   64.2$\pm$   2.0\\
J1435+3604&68\_12&   54.7$\pm$   6.7&...&   47.6$\pm$   2.9&...&   22.4$\pm$   2.8&  311.9$\pm$   3.4&  133.7$\pm$   4.1\\
J1437+5045&317\_38&  138.1$\pm$   1.9&    8.4$\pm$   1.3&   37.8$\pm$   1.1&   12.2$\pm$   1.0&   46.1$\pm$   1.2&  149.7$\pm$   1.1&   43.8$\pm$   0.8\\
J1445+3428&232\_33&   42.1$\pm$   1.2&    7.5$\pm$   0.9&   19.0$\pm$   0.8&    2.6$\pm$   0.6&   13.8$\pm$   0.8&   86.5$\pm$   1.0&   48.0$\pm$   0.8\\
J1514+3619&287\_14&   27.1$\pm$   1.3&...&    6.6$\pm$   2.7&...&    7.1$\pm$   1.3&   38.4$\pm$   0.9&   13.8$\pm$   1.3\\
J1550+4001&197\_23&$<$    4.5&...&$<$    2.5&...&...&$<$    2.6&...\\
J1555+3628&88\_11&  195.0$\pm$   2.1&   19.3$\pm$   1.4&   83.2$\pm$   1.9&...&   69.0$\pm$   1.8&  308.2$\pm$   1.9&  127.2$\pm$   1.8\\
J1616+4154&327\_30&  181.9$\pm$   5.6&...&   41.0$\pm$   2.9&   31.0$\pm$   2.8&   95.2$\pm$   3.0&  162.5$\pm$   2.4&   16.7$\pm$   1.5\\
J1617+0638&253\_39&$<$   25.2&...&$<$   17.6&...&...&$<$   18.0&...\\
J1619+3342&113\_40&   78.6$\pm$   2.1&...&   24.5$\pm$   1.9&...&   16.3$\pm$   2.4&  112.3$\pm$   2.0&   42.8$\pm$   1.9\\
J2257+1340&270\_40&$<$    2.4&...&$<$    3.2&...&...&$<$    2.6&   15.5$\pm$   1.2\\
J2345-0059&356\_12&$<$    3.1&...&$<$    3.4&...&    3.3$\pm$   1.0&...&...\\
\hline
&&&&Bonus: \\
\hline
J0820+2334&242\_9&   17.7$\pm$   1.0&...&    3.8$\pm$   0.6&...&    3.2$\pm$   0.6&   16.1$\pm$   0.5&    4.9$\pm$   0.5\\
J0910+1014&242\_34&$<$    7.9&...&$<$    6.7&...&...&$<$   25.7&...\\
J0910+1014&34\_46&  451.3$\pm$   4.4&   41.4$\pm$   2.6&  171.7$\pm$   4.3&...&  174.9$\pm$   3.9&  885.5$\pm$   3.7&  307.3$\pm$   3.6\\
J0914+2823&41\_123&   77.4$\pm$   5.0&...&   23.4$\pm$   3.7&...&   26.7$\pm$   3.0&  127.2$\pm$   5.0&   75.0$\pm$   4.4\\
J0925+4004&196\_22&$<$   11.6&...&$<$   14.8&...&...&...&...\\
J0928+6025&129\_19&  137.4$\pm$   5.0&...&   41.9$\pm$   3.8&...&   29.3$\pm$   3.6&  195.3$\pm$   2.6&   80.5$\pm$   2.0*\\
J0928+6025&187\_15&   80.1$\pm$   4.3&...&...&...&   28.6$\pm$   6.4&   95.8$\pm$   4.5&   10.9$\pm$   3.3\\
J0928+6025&188\_7&$<$   11.7&...&$<$    4.0&...&...&$<$    4.2&...\\
J0928+6025&90\_15&   48.7$\pm$   0.7&...&   20.2$\pm$   0.6&    3.5$\pm$   0.4&   12.6$\pm$   0.6&   70.7$\pm$   0.7&   28.6$\pm$   0.6\\
J0943+0531&216\_61&$<$    8.2&...&...&...&   18.4$\pm$   2.9&$<$    5.5&...\\
J0943+0531&227\_19&   26.7$\pm$   1.8&...&    5.4$\pm$   1.2&    3.2$\pm$   1.3&    6.0$\pm$   1.2&...&...\\
J0943+0531&29\_23&   23.7$\pm$   2.4&...&   14.7$\pm$   1.6&    7.7$\pm$   1.4&   22.4$\pm$   2.0&...&...\\
J1009+0713&170\_9&   77.1$\pm$   1.0&...&   34.1$\pm$   0.7&   20.8$\pm$   0.7&   61.2$\pm$   0.8&...&...\\
J1009+0713&86\_4&   12.7$\pm$   0.6&...&    3.5$\pm$   0.3&    3.0$\pm$   0.3&    6.8$\pm$   0.3&...&...\\
J1016+4706&359\_16&   80.7$\pm$   2.1&    8.4$\pm$   1.8&   37.9$\pm$   2.4&...&    9.7$\pm$   1.5&  138.9$\pm$   1.7&   75.8$\pm$   1.5\\
J1133+0327&164\_21&  116.0$\pm$   1.6&   12.2$\pm$   1.0&   36.7$\pm$   2.4&   11.6$\pm$   1.4&   21.6$\pm$   1.3&  155.1$\pm$   1.1&   40.2$\pm$   0.7*\\
J1133+0327&203\_10&$<$    3.8&...&$<$    3.5&...&...&$<$    2.4&...\\
J1233+4758&94\_38&  114.1$\pm$   2.3&   18.8$\pm$   1.5&   82.6$\pm$   3.1&   12.3$\pm$   2.0&   40.3$\pm$   2.3&  279.5$\pm$   3.4&   93.8$\pm$   2.2\\
J1233-0031&168\_7&   20.0$\pm$   1.4&...&    6.3$\pm$   1.1&...&    4.1$\pm$   1.2&   33.5$\pm$   2.5&    8.6$\pm$   1.2\\
J1241+5721&208\_27&   42.9$\pm$   2.5&    6.5$\pm$   1.5&   14.9$\pm$   2.4&...&   11.2$\pm$   2.2&   55.8$\pm$   2.0&   17.6$\pm$   1.8\\
J1330+2813&83\_6&   51.8$\pm$   1.5&...&   21.0$\pm$   0.6&    9.2$\pm$   0.6&   28.0$\pm$   0.7&...&...\\
J1342-0053&304\_29&   40.8$\pm$   5.1&...&   79.1$\pm$   2.4&...&   15.0$\pm$   2.6&  334.0$\pm$   0.0&  164.5$\pm$   2.1*\\
J1342-0053&77\_10&$<$   11.2&...&$<$   12.6&...&...&$<$    9.8&...\\
J1435+3604&126\_21&   21.5$\pm$   1.3&    2.2$\pm$   0.8&    8.8$\pm$   0.8&...&    7.9$\pm$   0.7&   55.5$\pm$   1.3&   22.1$\pm$   0.8\\
J1437+5045&24\_13&   27.2$\pm$   1.3&    3.8$\pm$   1.0&   12.9$\pm$   1.2&...&    9.7$\pm$   1.0&   85.2$\pm$   0.9&   35.5$\pm$   0.9\\
J1445+3428&231\_6&    8.6$\pm$   0.7&...&    8.6$\pm$   0.4&...&    1.9$\pm$   0.5&...&...\\
J1550+4001&97\_33&   38.5$\pm$   2.7&...&   10.0$\pm$   1.0&...&   11.1$\pm$   0.9&   57.6$\pm$   1.9&   21.1$\pm$   1.7*\\
J2257+1340&230\_25&   47.3$\pm$   2.5&...&   11.7$\pm$   2.5&...&...&   73.0$\pm$   2.7&   70.6$\pm$   3.6\\
J2257+1340&238\_31&   25.2$\pm$   3.4&...&   12.6$\pm$   3.0&...&...&   75.8$\pm$   4.7&   42.9$\pm$   5.0\\
\hline
\hline
\end{tabular*}
\caption{ Non-reddening corrected emission-line fluxes  (1) SDSS Field Identifier
(2) Galaxy Identifier, where the first number is the position angle in degrees from the QSO
and the second number is the projected separation in arcseconds (impact parameter) from the QSO
(3) - (8) Line fluxes are in units 10$^{-17}$ ergs s$^{-1}$ cm$^{-2}$ \AA$^{-1}$ 
[NII] fluxes marked with an asterisk indicate that the [NII] $\lambda$6584 was corrupted or fell outside the wavelength range of LRISr. We instead report the line flux of [NII]$\lambda$ 6548 scaled by its intrinsic factor of 2.96 \citep{osterbrock89}. In cases of non-emission line spectra, we report 3$\sigma$ upper limits to the flux at the positions of [OII], H$\beta$ and H$\alpha$.  \label{tab:gallines}}  
\end{table*}

\begin{table*}[hpt]\centering \scriptsize
\begin{tabular*}{0.95\textwidth}{@{\extracolsep{\fill}}llcccccrrcc}
\hline
Field&ID & z & $\rho$ &E(B-V) & Log(M$_{*}$) &$u-r$& SFR&
SFR & Abun & Abun \\
 & & & kpc&Balmer&   & & Balmer &
[OII] & M91 & PP04\\

\hline
&&&&&Targets: \\
\hline
\hline
J0042-1037&358\_9&0.0950& 15& 0.23$\pm$0.018& 9.32& 1.57$\pm$0.348&  0.18$\pm$ 0.02&  0.30$\pm$ 0.08& 8.34& 8.29\\
J0226+0015&268\_22&0.2274& 76&...&10.58&$>$ 2.08&$<$ 0.04&$<$ 0.06&...&...\\
J0401-0540&67\_24&0.2197& 81& 0.10$\pm$0.021& 9.92& 1.25$\pm$0.323&  1.14$\pm$ 0.15&  1.41$\pm$ 0.40& 8.55& 8.68\\
J0803+4332&306\_20&0.2535& 75&...&11.09&$>$ 2.81&$<$ 0.06&$<$ 0.25&...&...\\
J0820+2334&260\_17&0.0949& 28&...& 9.51&$>$ 2.08&  0.05$\pm$ 0.01&  0.10$\pm$ 0.03&...& 8.94\\
J0910+1014&35\_14&0.2647& 54&...&10.60&$>$ 2.08&$<$ 0.14&$<$ 0.09&...&...\\
J0914+2823&41\_27&0.2443& 99& 0.23$\pm$0.018& 9.59& 1.24$\pm$0.158&  2.83$\pm$ 0.34&  3.23$\pm$ 0.91& 8.62&...\\
J0925+4004&193\_25&0.2467& 92& 0.00$\pm$0.022&10.39& 1.71$\pm$0.192&  0.86$\pm$ 0.15&  0.56$\pm$ 0.16& 8.81&...\\
J0928+6025&110\_35&0.1540& 89&...&10.56& 2.55$\pm$0.222&$<$ 0.03&$<$ 0.04&...&...\\
J0935+0204&15\_28&0.2623&108&...&10.78&$>$ 2.23&$<$ 0.10&$<$ 0.04&...&...\\
J0943+0531&106\_34&0.2284&119& 0.43$\pm$0.024&10.57& 2.24$\pm$0.268&  4.52$\pm$ 0.58&  2.86$\pm$ 0.83& 8.89& 8.94\\
J0950+4831&177\_27&0.2119& 89&...&10.99& 2.74$\pm$0.273&$<$ 0.06&$<$ 0.17&...&...\\
J1009+0713&204\_17&0.2278& 59& 0.52$\pm$0.026& 9.63& 1.39$\pm$0.267&  4.58$\pm$ 0.61&  8.93$\pm$ 2.62& 8.32& 8.35\\
J1016+4706&274\_6&0.2520& 22& 0.00$\pm$0.019& 9.99& 1.48$\pm$0.230&  0.53$\pm$ 0.06&  0.68$\pm$ 0.19& 8.62& 8.66\\
J1022+0132&337\_29&0.0744& 39&...& 8.84&$>$ 1.02&  0.06$\pm$ 0.01&  0.05$\pm$ 0.02&...&...\\
J1112+3539&236\_14&0.2467& 52& 0.64$\pm$0.030&10.09& 1.42$\pm$0.227&  5.68$\pm$ 0.80& 10.66$\pm$ 3.20& 8.48& 8.72\\
J1133+0327&110\_5&0.2367& 18&...&11.00& 2.38$\pm$0.299&$<$ 0.29&$<$ 0.70&...&...\\
J1157-0022&230\_7&0.1638& 18&...& 2.27&$>$ 0.00&$<$ 0.09&$<$ 0.02&...&...\\
J1220+3853&225\_38&0.2737&152&...&10.53&$>$ 2.10&$<$ 0.06&$<$ 0.22&...&...\\
J1233+4758&50\_39&0.3826&195&...&10.90& 1.39$\pm$0.097&$<$ 0.53&$<$ 2.76&...&...\\
J1233-0031&242\_15&0.4714& 85&...&10.39&$>$ 1.54&  2.45$\pm$ 0.39&  2.35$\pm$ 0.68& 8.71&...\\
J1241+5721&199\_6&0.2053& 19& 0.80$\pm$0.037& 9.94& 1.42$\pm$0.167&  4.32$\pm$ 0.69& 12.45$\pm$ 3.93& 8.78& 8.54\\
J1245+3356&236\_36&0.1925&110& 0.12$\pm$0.022& 9.61& 1.27$\pm$0.170&  1.05$\pm$ 0.14&  1.15$\pm$ 0.33& 8.58& 8.36\\
J1322+4645&349\_11&0.2142& 36& 0.01$\pm$0.022&10.57& 2.02$\pm$0.241&  0.62$\pm$ 0.09&  0.19$\pm$ 0.06& 8.91&$>$ 8.95\\
J1330+2813&289\_28&0.1924& 85& 0.51$\pm$0.019&10.10&$>$ 2.50&  1.99$\pm$ 0.23&  2.38$\pm$ 0.67& 8.61&$>$ 8.95\\
J1342-0053&157\_10&0.2270& 34& 0.40$\pm$0.020&10.71& 1.56$\pm$0.059&  6.04$\pm$ 0.74&  1.35$\pm$ 0.38& 9.05& 8.71\\
J1419+4207&132\_30&0.1792& 87& 0.89$\pm$0.024&10.39& 1.75$\pm$0.132& 11.36$\pm$ 1.46& 14.21$\pm$ 4.13& 8.63&...\\
J1435+3604&68\_12&0.2024& 38& 0.84$\pm$0.020&10.87& 1.78$\pm$0.120& 18.96$\pm$ 2.28& 15.57$\pm$ 4.43& 8.81& 8.77\\
J1437+5045&317\_38&0.2460&140& 0.33$\pm$0.018& 9.92& 0.98$\pm$0.103&  4.29$\pm$ 0.50&  6.48$\pm$ 1.82& 8.48& 8.60\\
J1445+3428&232\_33&0.2176&111& 0.47$\pm$0.019&10.18& 1.92$\pm$0.299&  2.60$\pm$ 0.31&  2.76$\pm$ 0.78& 8.69& 8.92\\
J1514+3619&287\_14&0.2122& 46& 0.72$\pm$0.072& 9.46&$>$ 1.49&  1.96$\pm$ 0.51&  5.04$\pm$ 2.10& 8.68& 8.69\\
J1550+4001&197\_23&0.3125&101&...&11.11& 2.02$\pm$0.323&$<$ 0.06&$<$ 0.41&...&...\\
J1555+3628&88\_11&0.1893& 33& 0.26$\pm$0.018&10.31& 1.43$\pm$0.078&  4.18$\pm$ 0.49&  3.77$\pm$ 1.06& 8.74& 8.76\\
J1616+4154&327\_30&0.1036& 54& 0.33$\pm$0.021& 8.98& 1.08$\pm$0.148&  0.70$\pm$ 0.09&  1.28$\pm$ 0.37& 8.28& 8.29\\
J1617+0638&253\_39&0.1526& 99&...&11.30& 2.73$\pm$0.115&$<$ 0.08&$<$ 0.10&...&...\\
J1619+3342&113\_40&0.1414& 95& 0.48$\pm$0.022& 9.89& 1.67$\pm$0.181&  1.33$\pm$ 0.17&  2.07$\pm$ 0.59& 8.49& 8.72\\
J2257+1340&270\_40&0.1768&114&...&10.68& 2.80$\pm$0.319&$<$ 0.02&$<$ 0.14&...&...\\
J2345-0059&356\_12&0.2539& 45&...&10.61& 1.80$\pm$0.234&$<$ 0.14&$<$ 0.35&...&...\\

\hline
&&&&&Bonus: \\
\hline
J0820+2334&242\_9&0.0951& 15& 0.39$\pm$0.031& 8.95&$>$ 1.77&  0.07$\pm$ 0.01&  0.13$\pm$ 0.04& 8.51& 8.61\\
J0910+1014&242\_34&0.2641&132&...&11.22&$>$ 2.44&$<$ 0.30&$<$ 0.10&...&...\\
J0910+1014&34\_46&0.1427&110& 0.60$\pm$0.018&10.39& 1.51$\pm$0.074& 14.12$\pm$ 1.64& 20.46$\pm$ 5.75& 8.52& 8.67\\
J0914+2823&41\_123&0.2265&428& 0.65$\pm$0.033&10.59& 2.00$\pm$0.178&  6.42$\pm$ 0.96& 12.43$\pm$ 3.81& 8.46&$>$ 8.95\\
J0925+4004&196\_22&0.2475& 81&...&11.07&$>$ 2.58&$<$ 0.57&$<$ 0.45&...&...\\
J0928+6025&129\_19&0.1542& 48& 0.49$\pm$0.023& 9.86& 1.51$\pm$0.137&  2.89$\pm$ 0.37&  4.67$\pm$ 1.35& 8.47& 8.76\\
J0928+6025&187\_15&0.1537& 38&...& 9.33& 1.33$\pm$0.231&  0.45$\pm$ 0.05&  0.31$\pm$ 0.09&...& 8.31\\
J0928+6025&188\_7&0.2963& 29&...&10.80&$>$ 2.20&$<$ 0.08&$<$ 0.20&...&...\\
J0928+6025&90\_15&0.2931& 63& 0.21$\pm$0.018&10.03& 1.64$\pm$0.347&  2.27$\pm$ 0.27&  1.99$\pm$ 0.56& 8.77& 8.75\\
J0943+0531&216\_61&0.1431&147&...&10.73& 2.82$\pm$0.163&$<$ 0.02&$<$ 0.03&...&...\\
J0943+0531&227\_19&0.3530& 90&...& 9.37&$>$ 1.17&  0.47$\pm$ 0.11&  0.68$\pm$ 0.19& 8.49&...\\
J0943+0531&29\_23&0.5480&141&...& 9.80&$>$ 0.96&  3.67$\pm$ 0.55&  1.71$\pm$ 0.49& 8.82&...\\
J1009+0713&170\_9&0.3557& 43&...&10.02& 0.93$\pm$0.211&  3.04$\pm$ 0.31&  2.00$\pm$ 0.54& 8.73&...\\
J1009+0713&86\_4&0.3556& 19&...& 2.27&$>$ 0.00&  0.31$\pm$ 0.04&  0.33$\pm$ 0.09& 8.56&...\\
J1016+4706&359\_16&0.1661& 43& 0.25$\pm$0.020&10.26& 1.80$\pm$0.101&  1.37$\pm$ 0.17&  1.11$\pm$ 0.32& 8.83& 8.92\\
J1133+0327&164\_21&0.1545& 53& 0.39$\pm$0.021& 9.86& 1.29$\pm$0.143&  1.83$\pm$ 0.22&  2.57$\pm$ 0.73& 8.55& 8.55\\
J1133+0327&203\_10&0.2364& 35&...&10.66&$>$ 2.94&$<$ 0.03&$<$ 0.04&...&...\\
J1233+4758&94\_38&0.2221&130& 0.17$\pm$0.018&10.54& 2.13$\pm$0.192&  4.38$\pm$ 0.52&  2.11$\pm$ 0.59& 8.94& 8.66\\
J1233-0031&168\_7&0.3185& 31& 0.62$\pm$0.037&10.31& 1.38$\pm$0.338&  3.42$\pm$ 0.54&  6.10$\pm$ 1.92& 8.42& 8.54\\
J1241+5721&208\_27&0.2178& 91& 0.27$\pm$0.033& 9.82& 1.53$\pm$0.279&  1.06$\pm$ 0.17&  1.19$\pm$ 0.36& 8.66& 8.63\\
J1330+2813&83\_6&0.4164& 31&...&10.24&$>$ 1.57&  2.71$\pm$ 0.28&  1.95$\pm$ 0.53& 8.75&...\\
J1342-0053&304\_29&0.0708& 37& 0.39$\pm$0.018& 9.69& 1.87$\pm$0.076&  0.74$\pm$ 0.09&  0.17$\pm$ 0.05& 9.06& 8.86\\
J1342-0053&77\_10&0.2013& 31&...&10.28&$>$ 2.45&$<$ 0.08&$<$ 0.08&...&...\\
J1435+3604&126\_21&0.2623& 81& 0.80$\pm$0.023&10.15&$>$ 2.22&  5.56$\pm$ 0.70&  9.36$\pm$ 2.70& 8.44& 8.74\\
J1437+5045&24\_13&0.1430& 31& 0.85$\pm$0.024& 9.76&$>$ 2.22&  2.46$\pm$ 0.31&  3.76$\pm$ 1.09& 8.50& 8.76\\
J1445+3428&231\_6&0.6990& 41&...&11.19&$>$ 2.02&  3.86$\pm$ 0.43&  1.13$\pm$ 0.32& 9.04&...\\
J1550+4001&97\_33&0.3218&147& 0.71$\pm$0.024&10.68&$>$ 1.96&  7.41$\pm$ 0.96& 17.70$\pm$ 5.14& 8.62& 8.70\\
J2257+1340&230\_25&0.1781& 72& 0.79$\pm$0.041&10.53&$>$ 2.79&  2.96$\pm$ 0.50&  8.10$\pm$ 2.62&...&$>$ 8.95\\
J2257+1340&238\_31&0.1773& 89& 0.75$\pm$0.045&10.55&$>$ 2.20&  2.77$\pm$ 0.50&  3.59$\pm$ 1.20&...& 8.94\\

\hline
\hline
\end{tabular*}
\caption{ Derived Properties of target and bonus galaxies  (1) SDSS Field Identifier
(2) Galaxy Identifier, where the first number is the position angle in degrees from the QSO
and the second number is the projected separation in arcseconds (impact parameter) from the QSO
(3) Spectroscopic Redshift determined from zfind 
(4) Projected separation between the galaxy and QSO in kpc, calculated in the restframe of the galaxy.
(5) Balmer Correction and associated error, using intrinsic ratio of 2.86
(6) Stellar Mass from $kcorrect$ cite Blanton 
(7)  H$\alpha$ - derived star formation rate
(8) [OII]-derived star formation rate 
 (9) Oxygen Abundance from R23 according to the McGaugh 1991 Calibration 
(10)  Oxygen abundance from the N2 index of Pettini and Pagel 2004 \label{tab:galprops}}  
\end{table*}

\end{document}